\documentclass[pdftex]{pasj01}
\draft

\usepackage{multirow}
\usepackage[dvips]{graphicx}
\usepackage{pslatex}	    % to use PostScript fonts
\begin{document} 
\Received{}%{yyyy/mm/dd}
\Accepted{}%{yyyy/mm/dd}
%\Published{yyyy/mm/dd}

\title{Cloud-cloud collisions in the common foot point of molecular loops 1 and 2 in the Galactic Center}

%%% begin:list of authors
% Do NOT capitalize all letters in "textsc".
\author{Rei \textsc{enokiya}\altaffilmark{1}, Kazufumi \textsc{torii}\altaffilmark{2},  and Yasuo \textsc{fukui}\altaffilmark{1}}
%\thanks{Example: Present Address is xxxxxxxxxx}}
\altaffiltext{1}{Department of Physics, Nagoya University, Chikusa-ku, Nagoya, Aichi 464-8601, Japan}
\altaffiltext{2}{Nobeyama Radio Observatory, Minamimaki-mura, Minamisaku-gun, Nagano, 384-1305, Japan}
\email{enokiya@a.phys.nagoya-u.ac.jp}

%\author{B-Firstname \textsc{B-Familyname},\altaffilmark{2}}
%\altaffiltext{2}{B-Address of Institute}
%\email{bbbbb@xxx.xxx.xx.xx}

%\author{C-Firstname \textsc{C-Familyname}\altaffilmark{3}}
%\altaffiltext{3}{C-Address of Institute}
%\email{CCCcc@xxx.xxx.xx.xx}
%%% end:list of authors

%% `\KeyWords{}' always has to be placed before `\maketitle'.
\KeyWords{ISM: clouds --- ISM: kinematics and dynamics --- ISM: molecules --- stars: formation} %Do NOT move this preamble from here!

\maketitle

%How to write something
%1.7 \degree
%ra\fd ra\fm ra \fs
%dec\fh dec \fm dec \fs
%22 \micron

%*************************ABSTRACT HERE*********************************
\begin{abstract}
Recent large-area, deep CO surveys in the Galactic disk have revealed the formation of $\sim$50 high-mass stars or clusters triggered by cloud-cloud collisions (CCCs). Although the Galactic Center (GC) ---which contains the highest volume density of molecular gas--- is the most favorable place for cloud collisions, systematic studies of CCCs in that region are still untouched. Here we report for the first time evidence of CCCs in the common foot point of molecular loops 1 and 2 in the GC.
We have investigated the distribution of molecular gas toward the foot point by using a methodology for identifying CCCs, and we have discovered clear signatures of CCCs. Using the estimated displacements and relative velocities of the clouds, we find the elapsed time since the beginnings of the collisions to be 10$^{5-6}$ yr.
We consider possible origins for previously reported peculiar velocity features in the foot point and discuss star formation triggered by CCCs in the GC.
\end{abstract}

%*************************INTRODUCTION HERE*********************************
\section{Introduction}\label{intro}
\subsection{Cloud-cloud collisions and the Galactic Center}
 Due to the rapid compression of the molecular gas, a cloud-cloud collision (CCC) is a triggering mechanism for the formation of massive clumps that may become high-mass stars or a star cluster. \citet{hab92} performed axisymmetric numerical hydrodynamics calculations of supersonic, head-on collisions between non-identical clouds, and they showed that a collision between two molecular clouds with supersonic relative velocities generates a compressed layer at the collision front. \citet{ino13} pointed out that enhanced turbulence in the compressed layer increases the effective sound speed and consequently the effective Jeans mass. Recent large-area, deep CO surveys of the Galactic disk have revealed $\sim$50 objects containing high-mass stars that were formed via rapid gas compression by a CCC (e.g., super star clusters: \cite{fur09}; \cite{fuk13}, Galactic open clusters: \cite{tor11}; \cite{eno18}). Also, in extreme cases CCCs can trigger star bursts like those seen in mergers such as the Antennae \citep{wil00}, although the detailed mechanism by which this occurs is still unclear.

 The central kpc of the Galaxy (i.e., the Galactic Center: GC) contains $\sim$10 \% of the entire molecular contents of the Milky Way, and hence it is natural to expect clouds in the GC to collide with each other more frequently than in the Galactic disk. To elucidate the differences and coincidences between normal CCCs seen in the disk and the extreme CCCs seen as a star burst, a systematic analysis to probe for evidence of CCCs in the GC is strongly desired.
 The molecular complex in the central 200 pc, called the Central Molecular Zone (CMZ: \cite{mor96}), contains the highest volume density of molecular gas in the Galaxy; that is to say, it  appears to be the most favorable place for CCC to occur. Indeed, \citet{has94} discovered evidence for a CCC in the Sgr~B2 molecular complex. However, the velocity structure of the CMZ is generally fairly complicated, and it is quite difficult to distinguish colliding features from other velocity structures, such as high velocity compact clouds (\cite{oka98, tak19}). For this reason, the CMZ is not the most suitable region to explore for CCCs in the GC as the very first step.
 
 \subsection{Magnetic flotation loops in the GC}
There are four noteworthy CO condensations with large velocity dispersions in the GC in addition to the CMZ; namely, Bania's clumps 1 and 2, the L5.5 clump, and the common foot point of loops 1 and 2 (\cite{ban77, bit97, fuk06}). The nature of most of these massive clumps is still under debate. Only the origin of the third-brightest clump in the CO intensity map, ---the common foot point of loops 1 and 2, also known as M-3.8+0.9 (hereafter the foot point MC)--- is well understood; see \citet{fuk06} and subsequent observational and theoretical papers. 

\citet{fuk06} discovered two huge molecular loops with a projected length of $\sim$300 pc that connect with each other, and they proposed a magnetic flotation scenario due to the Parker instability \citep{par66} seen in solar prominences, based on its linear distribution in the $l-v$ diagram. \citet{tor10a} carried out higher-resolution follow-up observations toward the foot point MC using Mopra, Aacama Submillimeter Telescope Experiment (ASTE), and NANTEN2. They discovered not only the main component with $V_{LSR}$ ~$<$ -50 km~s$^{-1}$ ~which was originally identified by \citet{fuk06} as the foot point, but also a sub-component with $V_{LSR}$ ~$\ge$ ~-50 km~s$^{-1}$ that interacts with the main component. These two components form U shapes (U shape 1 and 2) in the $v-b$ diagram (Figure \ref{vb}). The investigators also discovered broad emission regions that display compact and broad velocity features with a very high $^{12}$CO intensity ratio between $J$=3--2 and $J$=1--0 and that bridge both sides of U shape 1 in the $v-b$ diagram. They concluded that the broad emissions originate from heating either by magnetic reconnection or by upward-flowing gas that bounced from the narrow neck at the foot point.
 \citet{tor10b} and \citet{kud11} carried out higher-$J$ CO observations and, by adopting a simple rotation-expansion kinematic model, they found that the observed, U-shaped velocity feature corresponds to the spatial U shape of the gas at the point connecting loops 1 and 2.
From observations of SiO($J$=2--1) and from mid-infrared spectroscopy, respectively, \citet{riq10} and \citet{kan12} discovered possible evidence for shock heating at the foot point, as originally proposed by \citet{fuk06}. Recently, through 3 mm line observations from Mopra and APEX, \citet{riq18} discovered the existence of slow, C-type shock waves with velocities of 30 -- 50 km~s$^{-1}$ ~at the foot point.

Some numerical magnetohydrodynamics (MHD) simulations have successfully reproduced the magnetic loops induced by the Parker instability (in two-dimensions, \cite{tak09}; in three-dimensions: \cite{mac09, suz15, kak18}).
 It is natural to expect clouds sliding down along the magnetic field lines from the top of a loop to stagnate at the foot points, thus increasing the chance of a CCC occurring there. From their numerical simulations, \citet{mat88} and \citet{tak09} found that shock waves are formed near the foot points of the magnetic loops where the supersonically infalling gas hits the dense disk gas. 
Motivated by these considerations, we have investigated the occurrence of CCCs in the foot point MC by applying our recently developed CCC-identification methodology in order to link our understanding of CCCs, as obtained from observations of the Galactic disk, to the GC.

\subsection{Observational Signatures of Colliding Clouds}
 \citet{fuk18a} have presented a detailed identification methodology for CCCs. Here, we focus only on the observational signatures of the colliding clouds. Assuming that the colliding clouds have different sizes ---because the chance of a collision between clouds of the same size is low--- the smaller cloud hollows out the larger cloud and makes a hole shaped like the smaller cloud in the larger cloud. This creates a complementary distribution between the two clouds at different velocities, unless the cloud is dispersed  by ionization originating from the triggered high-mass stars. If the angle between the direction of collision and the line of sight is zero degrees ($i.e$., for a head-on collision), the smaller cloud coincides with the hole in the larger cloud. In most cases, however, the angle is not zero, and the smaller cloud is displaced as a function of the angle and elapsed time relative to the hole in the larger cloud. In velocity space, the two clouds show a bridging feature at the beginning of the collision due to momentum exchange between them. This feature develops into a V shape in the $p-v$ diagram at a later stage of the collision. (For details, see \cite{fuk18a}.) This implies that the supersonic relative velocity of the two colliding clouds is converted into the velocity dispersion of the clouds via a CCC. Consequently, CCCs with large inter-cloud velocity dispersions ---as in the GC --- may form broad bridging features toward the colliding clouds. The most probable candidates are the broad emission regions in the foot point MC.

 By applying our CCC-identification methodology, in the present paper we provide for the first time evidence for CCCs in the foot point of the magnetic flotation loops in the GC. We adopt a distance of 8.0 kpc to the foot point MC. Details of the CO dataset we used are described in Section 2, and the results of our CO analyses are given in Section 3. We discuss the possible origins of CCCs, as well as differences or coincidences between CCCs in the disk and in the GC, in section 4. We summarize the present study in Section 5.
%*************** ABOVE: INTRODUCTION

%----------------- FROM HERE: Data
\section{Data}
\subsection{$^{12}$CO($J$=3--2)}
 In the present work, we have mainly used the $^{12}$CO($J$=3--2) ~dataset obtained by \citet{tor10a}. They carried out the observations in a position-switching mode with a 40\arcsec ~grid spacing at the $ASTE$ 10-m telescope of the NAOJ (National Astronomical Observatory of Japan) at Pampa La Bola at an altitude of 4800 m in Chile (\cite{koh05, eza04}; \yearcite{eza08}). The data coverage is limited roughly to ($l$, $b$) = (356\fdg0 to 356\fdg3, 0\fdg64 to 1\fdg27), corresponding to the location of the foot point MC. The HPBW (half-power beam width), the velocity resolution, and the $rms$ noise temperature in $T_\mathrm{mb}$ ~scale of the final data are 22\arcsec, 1.0 km~s$^{-1}$, and 0.29 K/km~s$^{-1}$, respectively.
\subsection{$^{12}$CO($J$=1--0)}
 The $^{12}$CO($J$=1--0) ~dataset we used was also obtained by \citet{tor10a}. The observations were carried out in the on-the-fly mode at the $Mopra$ 22-m telescope of the Australia Telescope National Facility near Coonabarabran in Australia. We used this data only to derive the masses and column densities of the molecular clouds, because the data coverage is slightly smaller and the angular resolution is coarser than that of the $^{12}$CO($J$=3--2) ~data. The HPBW, the velocity resolution, and the rms noise temperature in $T_\mathrm{mb}$ ~scale of the final data are 51\arcsec, 0.88 km~s$^{-1}$, and 0.3 K/ch, respectively.
%*************** ABOVE: DATA

%----------------- FROM HERE: RESULTS
\section{Results}
\subsection{Definition of Clouds from their Velocities}
 In the present section, we explore the possibility of CCCs between the main and the sub-components, focusing especially on the broad emissions.

\begin{figure}[!htbp]
  \begin{center}
    \includegraphics[width=11.5cm]{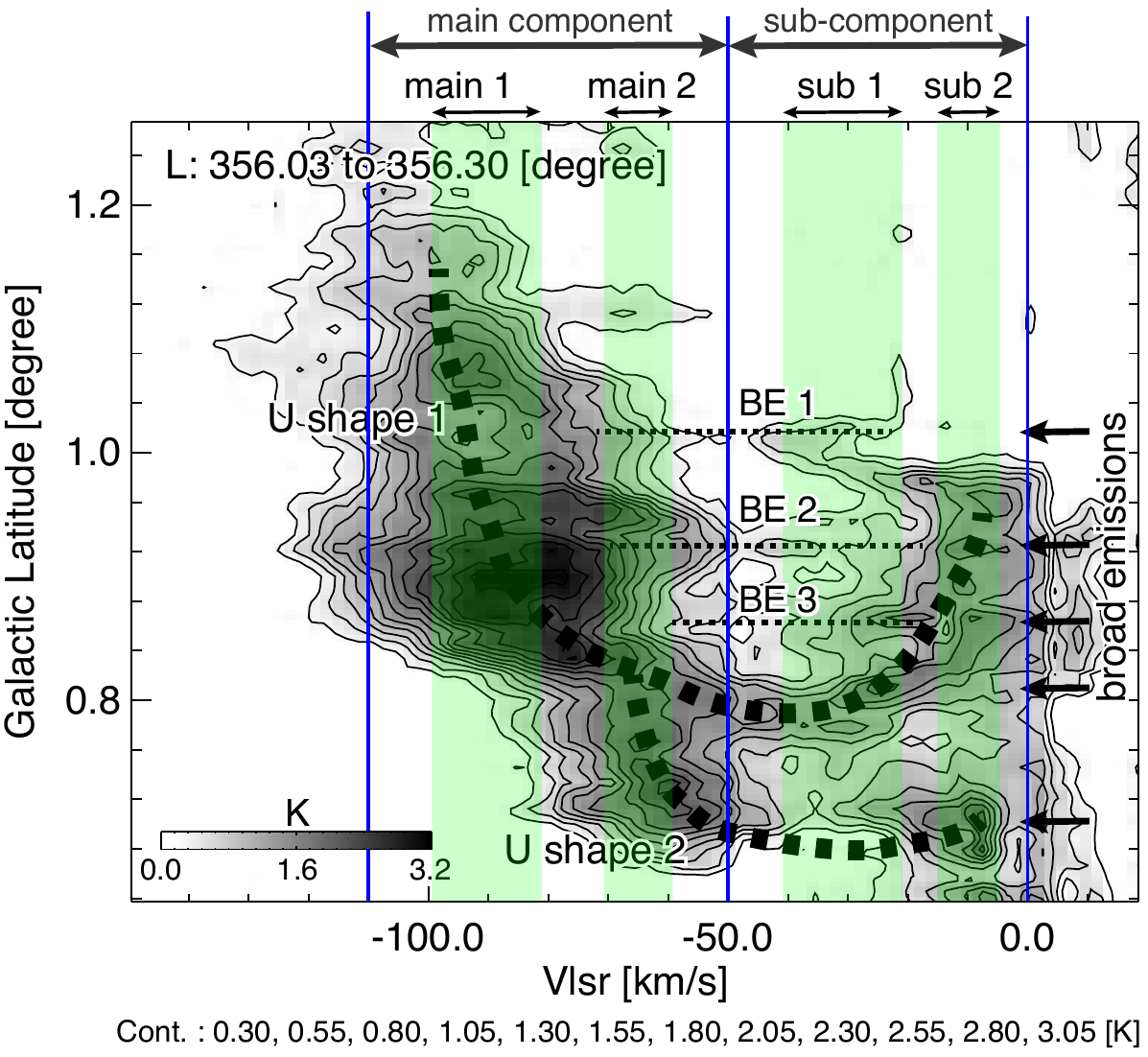}
  \end{center}
  \caption{Velocity-latitude diagram for $^{12}$CO($J$=3--2) ~in the longitude range from 356.03 to 356.30 degrees. The ranges of the velocity components and velocity features introduced by \citet{tor10a} are indicated by blue lines and black dashed lines respectively. The transparent green belts indicate the velocity ranges defined in Section 3.1.}\label{vb}
\end{figure}

Figure \ref{vb} shows the velocity--latitude diagram for $^{12}$CO($J$=3--2), with the range of integration extending over almost all longitudes. The velocity structures indicated by the thick, black, dashed lines ---corresponding to U shapes 1 and 2 ---and by the thin, black, dashed lines ---corresponding to the broad emission regions---were previously introduced by \citet{tor10a}. The two broad emission regions at the lowest latitude form U shapes 1 and 2, while the higher-latitude feature is a compact broad emission region. Here we call these three features broad emissions 1, 2, and 3 (indicated in the figure as BE 1, 2, and 3), from higher to lower latitude.
Fortunately, emission and absorption caused by foreground clouds along the line of sight are not seen significantly in this complex, except for absorption at $V_{LSR}$ = 0 to 5 km~s$^{-1}$.
As reviewed in Section \ref{intro}, a CCC is caused by an encounter between two or more clouds with different velocities. Consequently, it is necessary to determine how many velocity components the molecular complex has along the line of sight and how they are distributed.
The velocity ranges of the main component and the sub-component are quite large, and we expect that both contain multiple velocity components. To distinguish pairs of clouds with complementary spatial distributions in this complicated velocity structure, we defined the clouds according to their velocities, based on the distribution in the first-moment map \citep{fuk18a}.

\begin{figure}[!htbp]
  \begin{center}
    \includegraphics[width=17.5cm]{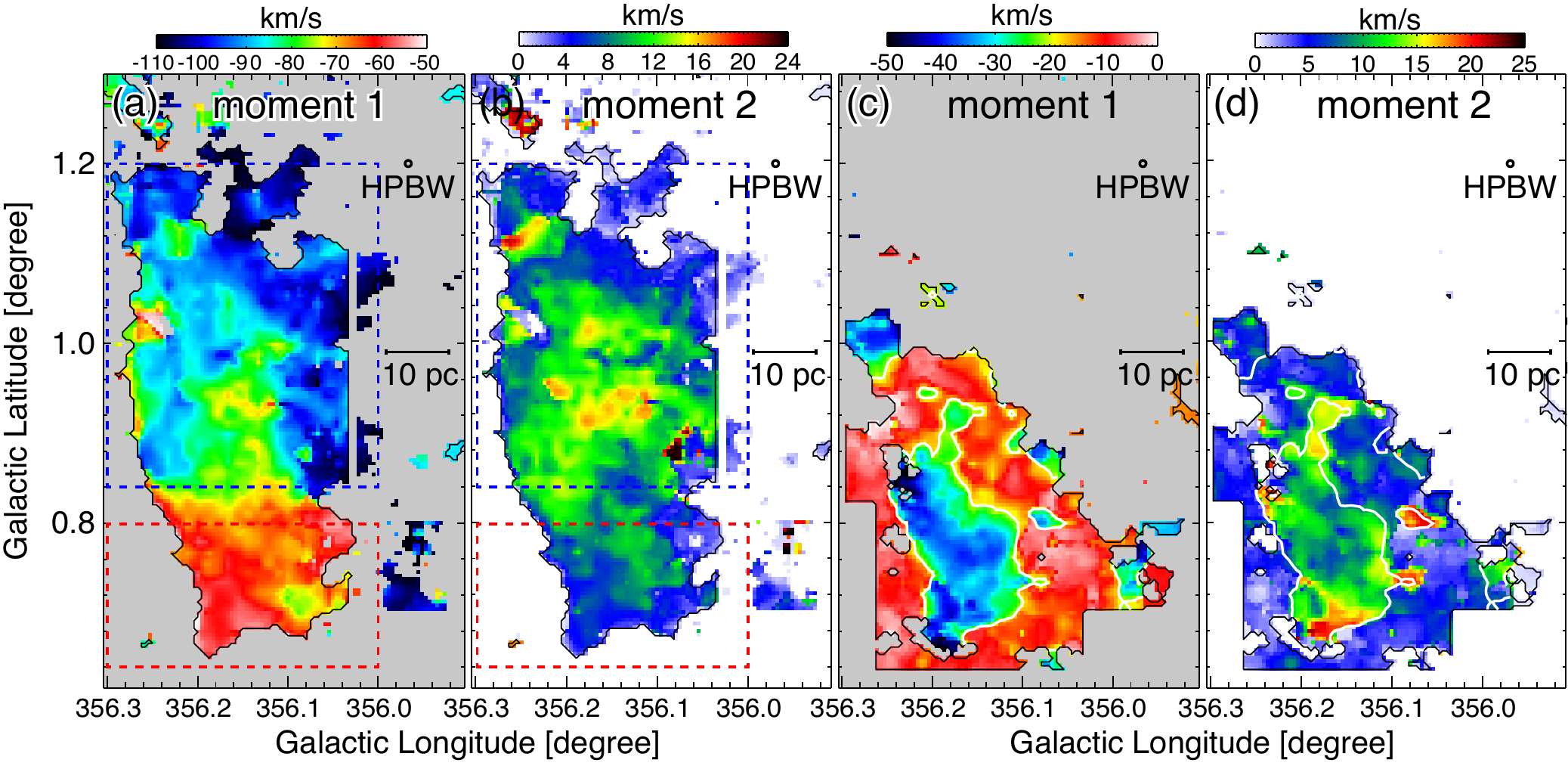}
  \end{center}
  \caption{(a, b) Intensity-weighted velocity map (first-moment map) and Intensity-weighted full-width-half-maximum map (second-moment map) of the main component in $^{12}$CO($J$=3--2). The blue and red dashed rectangles indicate the areas from which we obtained the values of V$_{center}$ for main component 1 and main component 2. (c, d) First-moment map and second-moment map of the sub-component in $^{12}$CO($J$=3--2). The white contours indicate the border lines for the calculation of V$_{center}$ for sub-component 1 and sub-component 2.}\label{mom}
\end{figure}
Figure \ref{mom}a and b show the first-moment and second-moment maps of $^{12}$CO($J$=3--2) ~in the velocity range of the main component. Voxels with intensities less than 4$\sigma$ are flagged in advance. Figure \ref{mom}a clearly shows the spatial distribution of the two components, the blue above $b$ $\sim$ 0\fdg8, and the red below $b$ $\sim$ 0\fdg8. We defined the region where the blue component dominates as ($l$, $b$) = (356\fdg00 to 356\fdg30, 0\fdg84 to 1\fdg20), as indicated by the dashed blue rectangle, and the region where the red component dominates as ($l$, $b$) = (356\fdg00 to 356\fdg30, 0\fdg64 to 0\fdg80), as indicated by the dashed red rectangle. By averaging the values inside each rectangle in the first-moment and the second moment maps, we obtain the central velocities ($V_\mathrm{center}$) and velocity widths ($dV$) for each component (see Table \ref{table:physpara}). Here, we call these blue-shifted and red-shifted components ``the main component 1'' and ``main component 2'', respectively.
Figure \ref{mom}c and d show the first-moment and second-moment maps of $^{12}$CO($J$=3--2) ~in the velocity range of the sub-component. Voxels with intensities less than 4$\sigma$ are flagged in advance. The sub-component is clearly separated in space in both the first- and second-moment maps. We defined the region dominated by the blue-shifted component as $<$ -20 km~s$^{-1}$ ~corresponding to the interior of the white contour, and the red-shifted component as $\ge$ -20 km~s$^{-1}$. By averaging the values for each region in the first-moment and second-moment maps, we obtained the central velocities ($V_\mathrm{center}$) and velocity line widths ($dV$) for each component (see Table \ref{table:physpara}). We call these blue-shifted and red-shifted components ``sub-component 1'' and ``sub-component 2'', respectively.
We also define the velocity-integration range where each component predominates as $V_\mathrm{center}$ $\pm$ $dV$, although the clouds are more extended in velocity and connect with each other kinematically to form the U shapes in the $v-b$ diagram (see Table \ref{table:physpara}).
Main component 1 includes the blue-shifted component of U shape 1, and main component 2 includes the blue-shifted component of the broad emissions and the blue-shifted component of U shape 2. sub-component 1 includes the red-shifted component of the broad emissions and the bottoms of the U shapes, and sub-component 2 includes the red-shifted components of U shapes 1 and 2.

\begin{figure}[!htbp]
  \begin{center}
    \includegraphics[width=17.5cm]{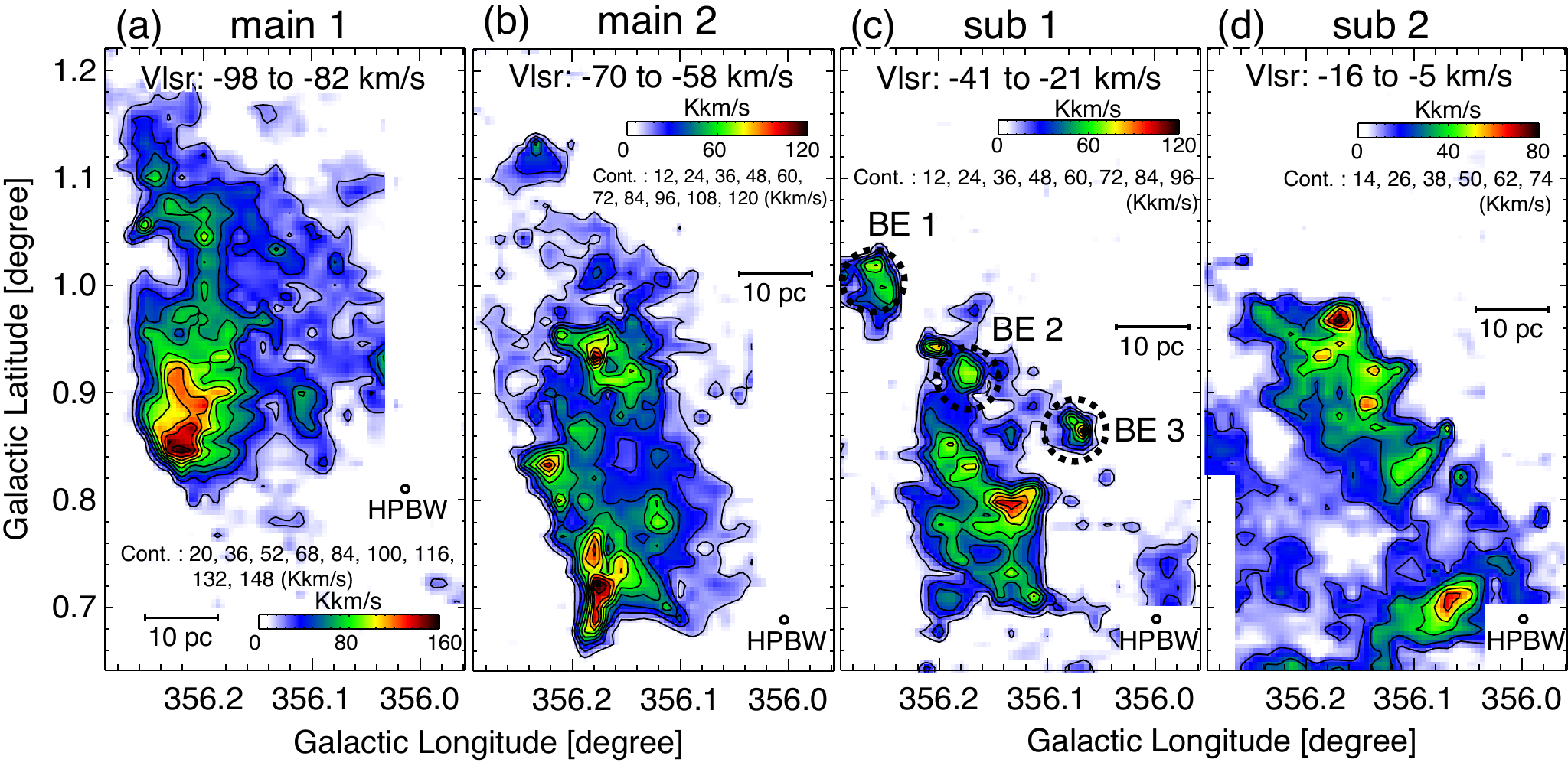}
  \end{center}
  \caption{(a-d) Integrated intensity distributions of main components 1 and 2 and sub-components 1 and 2 in $^{12}$CO($J$=3--2). The velocity-integration ranges are derived from the moment maps (see details in text).}\label{mcs}
\end{figure}

Figure \ref{mcs} a--d show the integrated-intensity distributions of main components 1 and 2 and sub-components 1 and 2. The velocity-integration range for each cloud corresponds to $V_\mathrm{center}$ $\pm$ $dV$. Main component 1 is shaped like Swiss cheese. It contains a huge $\sim$40 pc molecular complex from $b$ = 0\fdg8 to 1\fdg1, with three holes or depressions at ($l$, $b$) = (356\fdg24, 1\fdg02), (356\fdg15, 1\fdg01), and, (356\fdg15, 0\fdg94) (hereafter, holes 1, 2, and 3). Main component 2 exhibits a complicated morphology. The upper half comes from a mixture of main component 1 and sub-component 1. The lower half exhibits a concentration of molecular gas. sub-component 1 consists of the broad emission regions 1, 2, and 3 and a molecular complex at $b$ $< 0\fdg9$. The latter corresponds to the bottom of U shape 1 \citep{tor10a}. sub-component 2 exhibits a huge elliptical hole centered at ($l$, $b$) $\sim$ (356\fdg18, 0\fdg80), almost at the center of the molecular cloud. This hole can be seen more clearly at $V_{LSR}$ ~from -10 to 10 km~s$^{-1}$ ~in Figure \ref{lbch2} in the Appendix. 

To estimate the hydrogen mass and column density of each component, we utilize the following equations:

\begin{equation}
  M = \mu m_p \sum_{i} ~[d^2 \Omega N_{H_2, i}].
\end{equation}
where $\mu$, $m_p$, $d$, $\Omega$ and $N_\mathrm{H_2, i}$ are the mean molecular weight, proton mass, distance, solid angle subtended a pixel, and column density of molecular hydrogen for the i-th pixel, respectively. We assume a helium abundance of 20 $\%$, which corresponds to $\mu$ = 2.8, and we take $d$ = 8.0 kpc. The column density of molecular hydrogen is given by Equation (2),

\begin{equation}
  N_\mathrm{H_2} = X \times W(CO).
\end{equation}
where $W$(CO) is the integrated intensity of $^{12}$CO($J$=1--0) ~and $X$ is an empirical factor that converts from $W$(CO) to $N_\mathrm{H_2}$. We adopt $X$ = 0.7 $\times$ 10$^{20}$ cm$^{-2}$ (K km~s$^{-1}$)$^{-1}$, as obtained by \citep{tor10b} by comparing $IRAS$ dust-emission data and CO datasets toward loops 1 and 2. From these equations and the $^{12}$CO($J$=1--0) ~dataset obtained with Mopra, we found the molecular masses for main components 1 and 2 and sub-components 1 and 2 to be $\sim2\times10^5$ $M_\odot$, $\sim9\times10^4$ $M_\odot$, $\sim8\times10^4$ $M_\odot$, and $\sim8\times10^4$ $M_\odot$, respectively. The detailed physical parameters for each cloud are summarized in Table \ref{table:physpara}. We used the end-to-end velocity of each cloud, shown as $V_{LSR}$ ~in Table \ref{table:physpara}, to derive the mass and column density.

\begin{table*}[!htbp]
\tbl{Physical parameters of the molecular clouds}{%
\begin{tabular}{lccccc} 
\hline\noalign{\vskip3pt}
      cloud name & $V_{LSR}$ ~[km~s$^{-1}$] & $V_\mathrm{center}$ ~[km~s$^{-1}$]  & $dV$ ~[km~s$^{-1}$] & $N_\mathrm{H_2}$ ~(peak/mean) [$\times$10$^{22}$ cm$^{-2}$] & Mass [$\times$10$^{4}$ $M_\odot$] \\
\hline\noalign{\vskip3pt} 
    main component 1 & -110.0 -- -70.0 & -89.8 & 8.0 & 2.2 / 1.0 & 19.7 \\
    main component 2 &  -70.0 -- -50.0 & -64.1 & 5.9 & 1.4 / 0.5 & 9.2  \\
    sub-component 1 & -50.0 -- -20.0 & -31.1 & 10.1 & 1.1 / 0.4 & 8.3 \\
    sub-component 2 & -20.0 -- 0.0 & -10.2 & 5.3 & 0.9 / 0.4 & 8.3 \\
\hline\noalign{\vskip3pt} 
\end{tabular}}
\label{table:physpara}
\begin{tabnote}
\hangindent6pt\noindent
Note. --- Col.1: Names of components. Col.2: Velocity ranges. Col.3: Peak velocities derived from moment-1 maps. Col.4: Velocity line widths derived from moment-2 maps. Col.5: Maximum molecular column density toward each cloud. Col.6: Molecular mass.\\
\end{tabnote}
\end{table*}

%%%%%%%%%%%%%%%%%%%%%%%%%%%%%%%%%%%%%%%%%%%%%%%%
\subsection{Pairs of Complementary Distributions among the Molecular Components}
 According to \citet{fuk18a} and references therein, the signatures of a CCC are the complementary distributions in space of the colliding clouds and a broad bridging feature or a V-shaped feature connecting the clouds in velocity. To investigate evidence for CCCs in the foot point MC, we assess the complementary distributions among the identified components in this subsection.
We first take particular note of the broad features 1, 2, and 3, because they may be candidates for bridging features.
\begin{figure}[!htbp]
  \begin{center}
    \includegraphics[width=11.5cm]{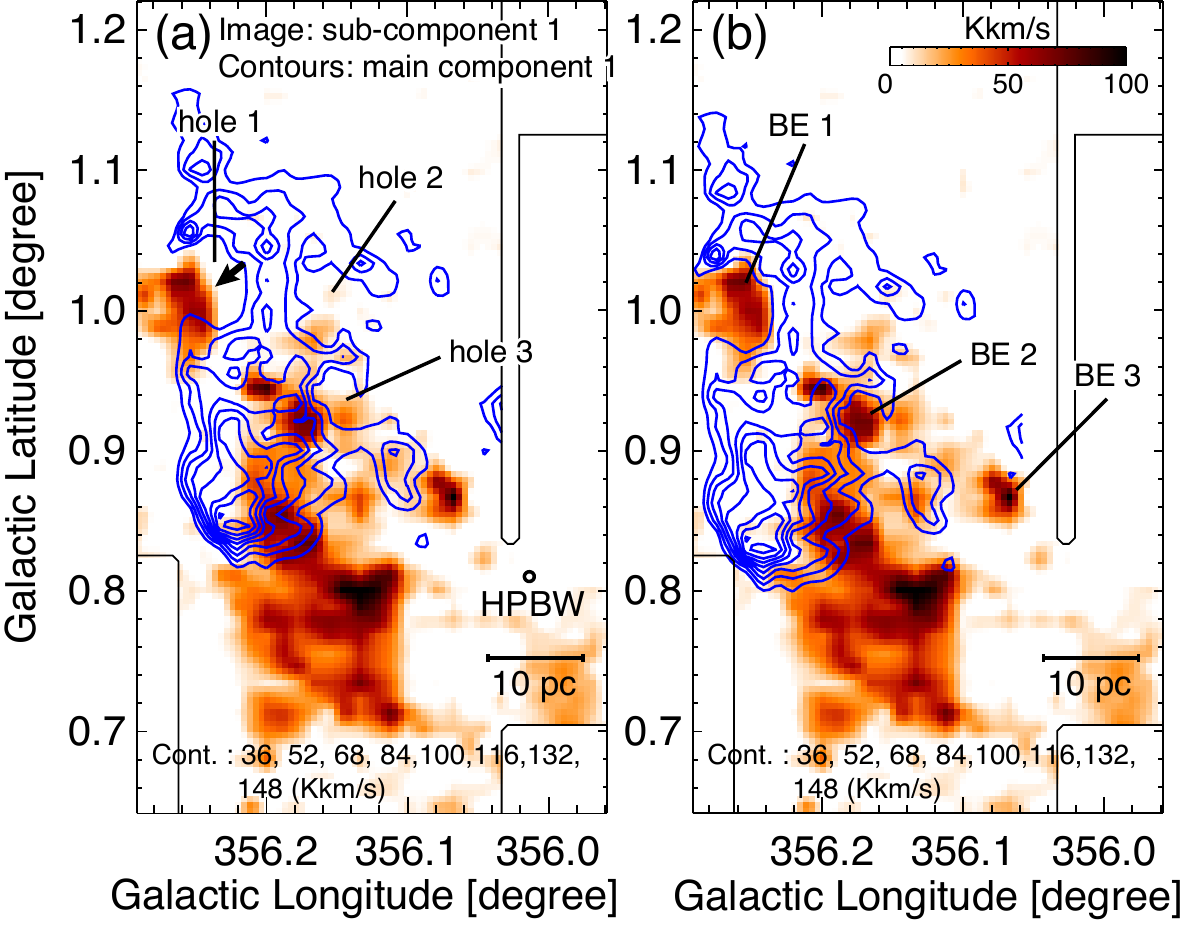}
  \end{center}
  \caption{(a) Integrated intensity distribution of sub-component 1 (image) and main component 1 (contours). The black arrow indicates the plausible displacement vector (see text). (b) The same as panel (a) but with the contours displaced by the vector.}\label{disp1}
\end{figure}

Figure \ref{disp1}a shows the integrated $^{12}$CO($J$=3--2) ~intensity distribution of sub-component 1 as contours superposed on a background image of main component 1.
We found that some of the holes in main component 1 (the image) seem to fit some of the broad emissions.
To estimate the relative displacement between these components, we applied an algorithm developed by Fujita et al. (2019 in preparation). This algorithm determine a plausible displacement vector by calculating correlation coefficients between the integrated intensity map of the hole cloud and that of the clump cloud with an arbitrary displacement. The displacement vector with the most negative correlation coefficient is consider to be the plausible displacement vector. By applying this algorithm, we found the magnitude and angle of the displacement between this pair of components to be 4.20 pc and 213.69 degrees, where a positive angle indicates counter-clockwise rotation from Galactic east.
The black arrow in Figure \ref{disp1}a represents this plausible displacement vector. 

Figure \ref{disp1}b is the same as Figure \ref{disp1}a, but with the positions of the contours shifted by the displacement vector. This figure clearly exhibits complete correspondence between broad emission 1 and hole 1, and between broad emission 2 and hole 3. This is clear evidence of CCCs. Moreover, the molecular complex at $b$ $< 0\fdg9$ also corresponds better with main component 1 in Figure \ref{disp1}b. Molecular gas in the vicinity of broad emission 3 is tenuous, and we did not find any significant pairs of cloud with a hole-shape in any other velocities, even though the emission connects main components 1 and 2 in velocity space, as shown by Figure\ref{vb}.

\begin{figure}[!htbp]
  \begin{center}
    \includegraphics[width=12.5cm]{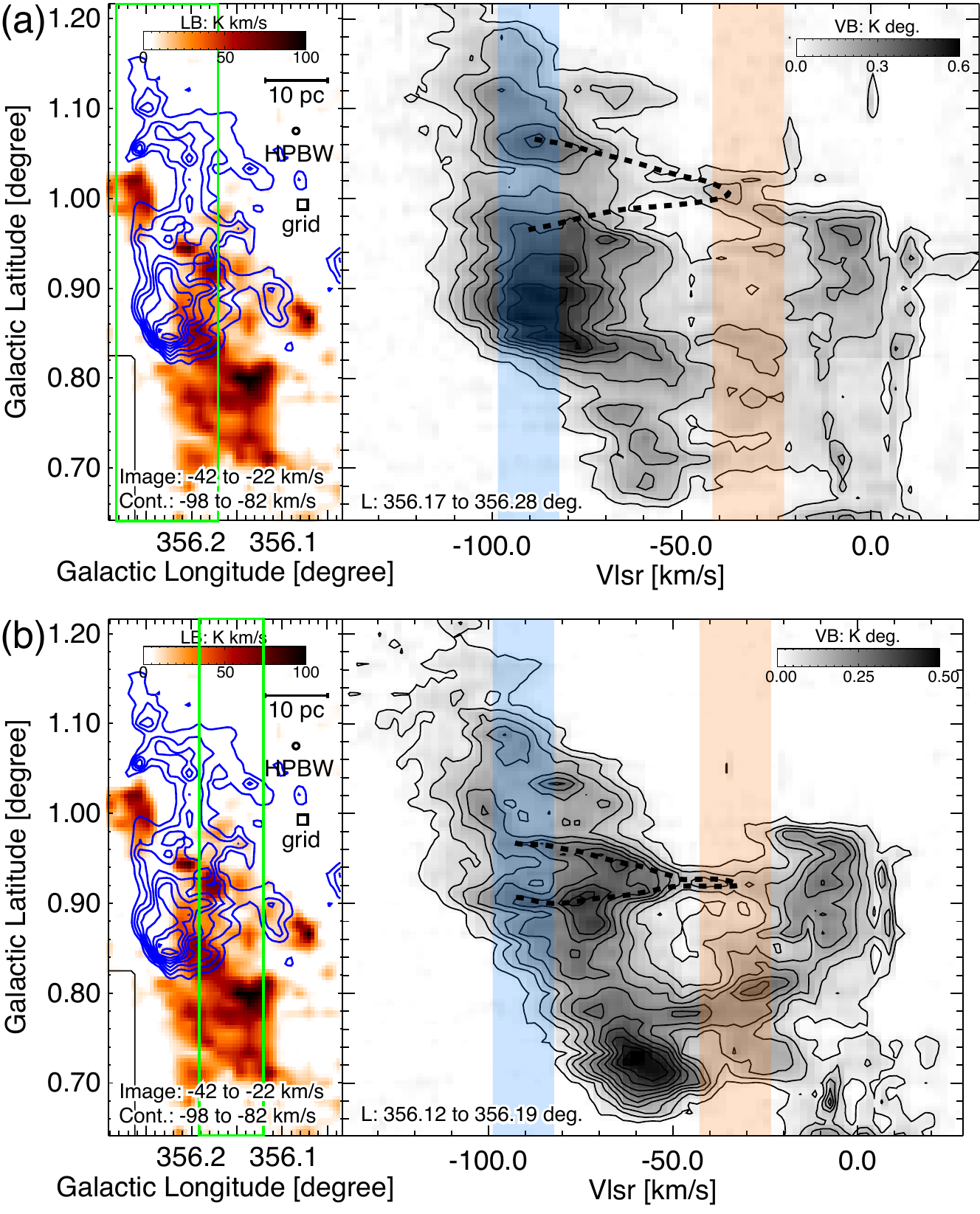}
  \end{center}
  \caption{(a) ($Top-left$) Integrated intensity distribution of sub-component 1 (color image) and main component 1 (blue contours) in $^{12}$CO($J$=3--2). ($Top-right$) $v-b$ diagram of $^{12}$CO($J$=3--2) ~toward broad emission 1. The blue and orange transparent belts indicate the respective velocity-integration ranges of the two components. The light green rectangle in the left panel indicates the longitude-integration range for the $v-b$ diagram. (b) The same as panel (a) but toward broad emission 2.}\label{lbv1}
\end{figure}

Next we examine the velocity structures toward the candidate collision sites. Figures \ref{lbv1}a and b show the longitude-latitude and velocity-latitude distributions toward broad emissions 1 and 2. The velocity-integration ranges for main component 1 and sub-component 1 are indicated by the transparent blue and orange belts in the $v-b$ diagram. The longitude-integration range is indicated by the light green rectangle in the $l-b$ map.
The figure obviously shows that both pairs of clouds make V shapes in the $v-b$ diagrams, as indicated by the thick, black, dashed lines. This is another signature of a CCC.

 Secondly, we investigate other pairs of colliding clouds in the foot point MC between the main and the sub-components.

\begin{figure}[!htbp]
  \begin{center}
    \includegraphics[width=11.5cm]{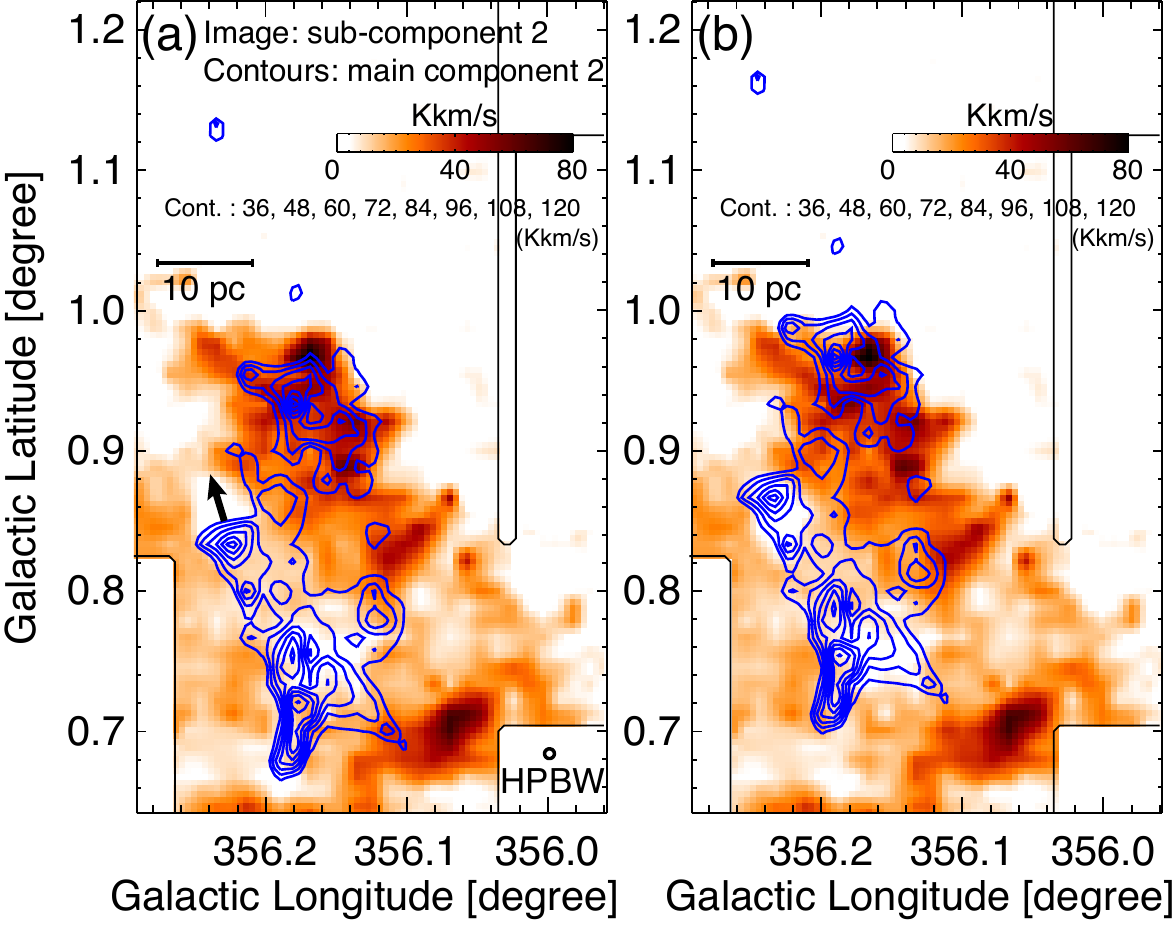}
  \end{center}
  \caption{(a) Integrated intensity distributions of sub-component 2 (image) and main component 2 (contours) in $^{12}$CO($J$=3--2). The black arrow indicates the plausible displacement vector. (b) The same as the panel (a), but with the contours displaced by the vector.}\label{disp2}
\end{figure}
 Figure \ref{disp2}a shows the blue contours of main component 2 superposed on the color image of sub-component 2 in $^{12}$CO($J$=3--2). The huge elliptical hole in sub-component 2 seems to fit the molecular concentration below $b$=0\fdg85 in main component 2. Applying the algorithm again, we obtain the magnitude and angle of the displacement of the pair of components to be 4.97 pc and 110.56 degrees. The black arrow in Figure \ref{disp2}a is the plausible displacement vector derived by the algorithm. Figure \ref{disp2}b shows the distribution shifted by this vector. In the velocity range of main component 2, emission from main component 1 and sub-component 1 are contaminated at $b > $ 0\fdg85, and we therefore excluded this region in deriving the displacement vector.

\begin{figure}[!htbp]
  \begin{center}
    \includegraphics[width=12.5cm]{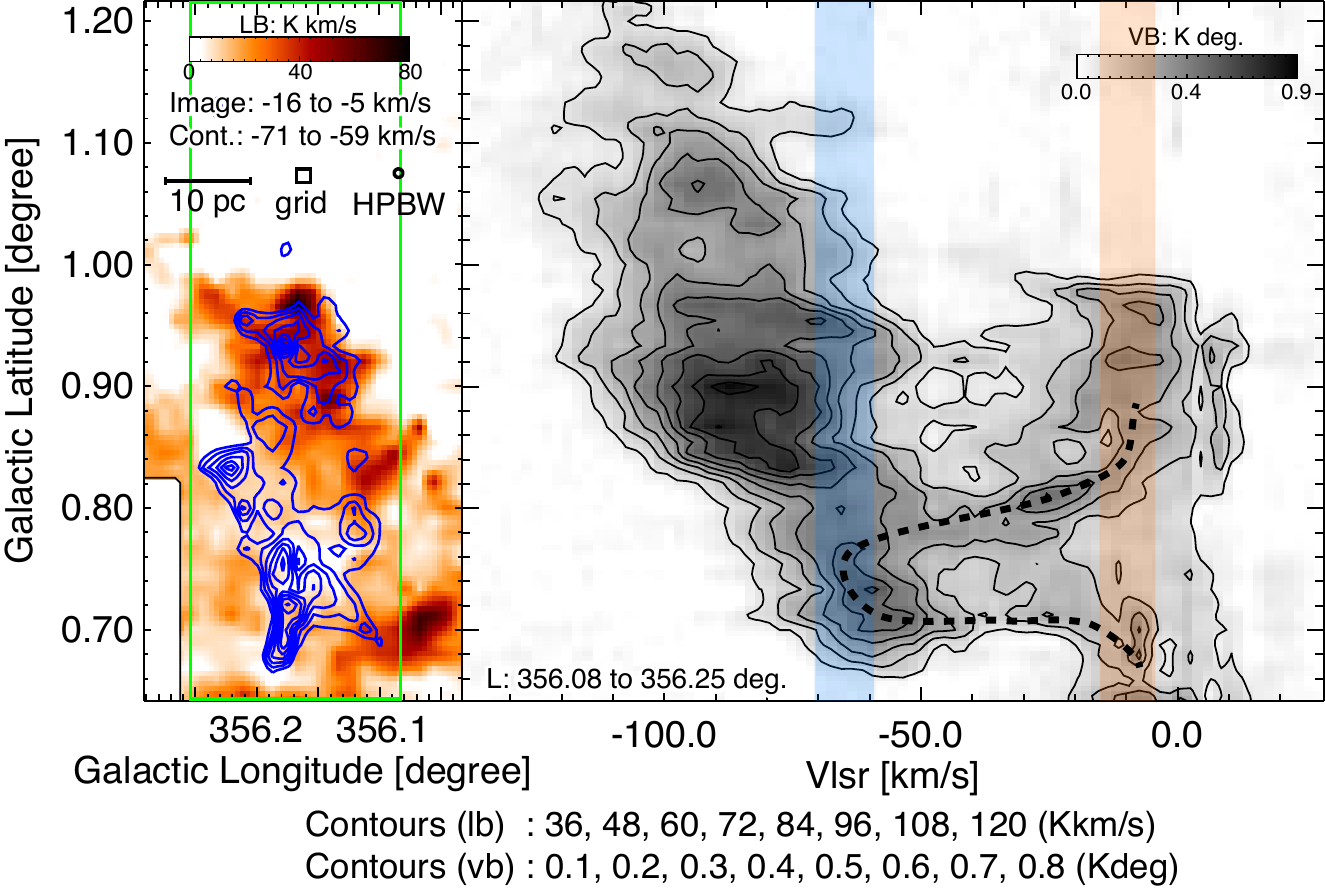}
  \end{center}
  \caption{($Left$) Integrated intensity distribution of sub-component 2 (color image) and main component 2 (blue contours) in $^{12}$CO($J$=3--2). ($Right$) $v-b$ diagram of $^{12}$CO($J$=3--2) ~toward the bottom of U shape 1. The blue and orange transparent belts indicate the respective velocity-integration ranges of the two components. The light green rectangle in the $l-b$ diagram indicates the longitude-integration range for the $v-b$ diagram.}\label{lbv2}
\end{figure}

Figure \ref{lbv2} shows the longitude-latitude, and velocity-latitude distributions of the pair of candidate CCC sites between main component 2 and sub-component 2. The light green rectangle indicates the longitude-integration range for the $v-b$ diagram, and the blue and orange transparent belts indicate the respective velocity ranges of main component 2 and sub-component 2. As indicated by the thick, black, dashed lines, the candidate clouds exhibit complementary spatial distributions, making a V shape in the $v-b$ diagram that is characteristic of a CCC. For this reason, it is possible that this pair of clouds is also colliding with each other. This V shape is constructed from U shape 1 and U shape 2 identified by \citep{tor10a}.
%*************** ABOVE: RESULTS

%----------------- FROM HERE: DISCUSSION
\section{Discussion}
In the previous section, we have shown that two pairs of velocity components (main component 1--sub-component 1, hereafter CCC1; and main component 2--sub-component 2, hereafter CCC2) exhibit complementary distributions in space and V shapes in the $v-b$ diagram. These are strong evidence for current CCCs. In this section, we discuss detailed scenarios for these CCCs as well as the differences or coincidences between CCCs in the Galactic disk and in the GC.

\subsection{Collision timescale}
 A spatial displacement between a hole cloud and a clump cloud in a CCC is produced by the product of the relative velocity toward the direction of the displacement vector and the elapsed time since the collision. 
Assuming the relative velocity in the direction of the displacement is the same as their relative velocity along the line of sight ---58.7 km~s$^{-1}$ ~and 53.9 km~s$^{-1}$--- $i.e$., that the angle between the line of sight and the collision direction is 45 degrees, we obtain the elapsed time ---or collision timescale--- to be 4.2 pc / 58.7 km~s$^{-1}$ ~= $\sim7 \times10^4$ years for CCC1 and 4.97 pc / 53.9 km~s$^{-1}$ ~= $\sim$9 $\times$ 10$^4$ years for CCC2.
This velocity assumption is commonly used for CCCs in the Galactic disk, because the collision is caused by random motions of clouds within the velocity dispersion of the spiral arm where the colliding clouds are located. On the other hand, the velocity of the foot point MC is dominated by systematic down-flow motions along the magnetic field lines in loops 1 and 2 to the foot point. Thus, the relative velocity toward the displacement vector depends strongly on the collision angle, and it is quite unclear whether or not the above assumption is applicable for this region. However, in any case the line width of each component is typically 20 -- 30 km~s$^{-1}$, and the Alfv\'en speed estimated by \citet{fuk06} is $\sim$24 km~s$^{-1}$, so we can expect the relative velocity in any direction to be more-or-less these values. For these reasons, we used 24 km~s$^{-1}$ ~for the relative velocity, and accordingly we estimate the collision timescale for both CCC1 and CCC2 to be $\sim$10$^{5-6}$ years.

This collision timescale is enough shorter than the several-million-year formation timescale of the loops \citep{fuk06} for this CCC scenario to be consistent with the magnetic loop model. Although the collision timescale is long enough for the collision to form high-mass star(s) in the compressed layer, \citet{tor10a} reported that there is no star-forming activity in the foot point MC.

\subsection{CCC Scenario}
\begin{figure}[!htbp]
  \begin{center}
    \includegraphics[width=16.5cm]{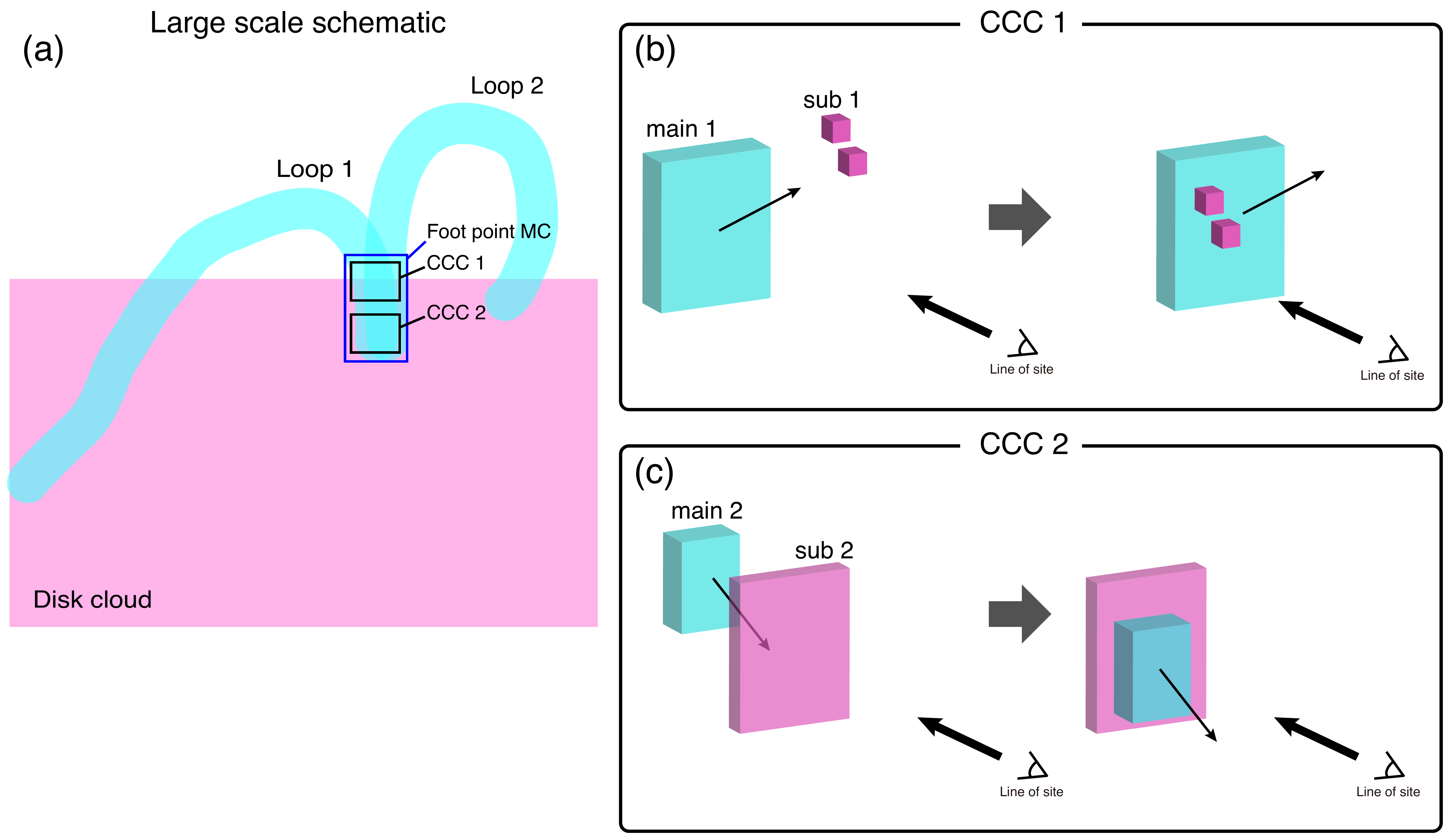}
  \end{center}
  \caption{(a) Schematic image of the region toward loops 1 and 2. (b, c) Schematic images of CCC1 and CCC2. The upper and lower panels show the configuration of the clouds prior to the collisions and at present, respectively. The main components and the sub-components are represented by cyan and pink, respectively.}\label{sche}
\end{figure}

The plausible displacement vectors indicate that main component 1 collided with sub-component 1 from Galactic southeast to the northwest (CCC1), while main component 2 collided with sub-component 2 from Galactic north to the south (CCC2). Figure \ref{sche}a shows a large-scale schematic image toward molecular loops 1, 2, and the foot point MC. Figures \ref{sche}b and c are close-up schematic images toward the two CCCs.

\citet{fuk06} first discovered the molecular loops and determined the velocity of the foot point to be $\sim$-130 to $\sim$-50 km~s$^{-1}$ ~in $^{12}$CO($J$=1--0). Figure 7 of \citet{tor10b} [the velocity-channel distribution of $^{12}$CO($J$=1--0)] also shows that loops 1, 2, and the foot point MC are only seen in this above velocity range (the main component). From $^{12}$CO($J$=3--2) ~observations with higher angular resolution, \citet{tor10a} discovered another velocity feature, the sub-component, that is interacting with the main component. According to these observations, they suggested that the main component consists mainly of loops 1 and 2 and that the sub-component consists of molecular clouds in the Galactic disk, $i.e$, that the molecular clouds were originally located at the foot point of loops 1 and 2 and did not float upward due to magnetic buoyancy. This indicates that the CCCs discovered in the present work are collisions between loop clouds and disk clouds.

On the other hand, \citet{tor10b} and \citet{kud11} proposed a simple dynamical model for the main component and the sub-component as loop 1 and loop 2, respectively. This model indicates that the CCCs are caused by contacts between loop 1 and loop 2. Although a detailed kinematic model of the foot point and the interpretation of the U shape are still under debate, 

we can interpret our CCC scenario through either of the two kinematic models. For a loop-disk collision, the loop clouds ---in other words, the main components--- might be falling down from Galactic north to the south to collide with the disk clouds. However, the CCC1 is inconsistent with this model. \citet{mat88} indicates that numerical simulations found shock waves near the foot points of the magnetic loops where the supersonically infalling gas hits the dense disk gas.
For a loop-loop collision, both loop clouds may be falling down with some relative velocity. In this case, both north-to-south and south-to-north collisions are acceptable. Accordingly, a loop-loop collision appears to be consistent with the kinematic model, even though a loop-disk collision is not completely ruled out. The high intensity ratio between $^{12}$CO($J$=3--2) ~and $^{12}$CO($J$=1--0) ~and the diffuse SiO emission throughout the foot point MC probably imply that both the loop-disk and loop-loop collisions took place elsewhere. Further detailed study of the kinematics in the foot point based on MHD simulations is required.

\subsection{The Origin of the Velocity Features in the foot point MC}
\citet{haw15} produced synthetic CO observations from numerical hydrodynamical simulations of a CCC obtained by \citet{tak14}. Figure 7 of \citet{haw15} clearly shows that the initial relative velocity between the two clouds is reduced during the collision via momentum exchange. This figure also shows that the large cloud (or hole cloud) maintains its initial velocity, whereas the velocity of the small cloud (or clump cloud) shifts to an intermediate velocity between the initial velocities of the two clouds.

Assuming this mechanism to be universal, we here propose an origin for the U shapes and broad emissions in the $v-b$ diagram. A possible scenario is as follows: The collision started 10$^{5-6}$ years ago. At this stage, the main component had only the one velocity feature corresponding to the velocity of the present main component 1, and the sub-component also had only the one velocity feature corresponding to the velocity of the present sub-component 2. Next, momentum exchange via CCCs created the intermediate velocity features ---namely, the present main component 2 and the present sub-component 1--- at the interacting parts of the two components. Clump clouds in the sub-component created the broad emissions observed in CCC1, and a clump cloud at the bottom of the main component created the bottom parts of U shape 1 and U shape 2 observed in CCC2. This proposed CCC-based interpretation of the $v-b$ diagram is consistent with the previous kinematic models of the molecular loops.

%%%revision%%%
\citet{tor10a} explained the origin of high temperatures of $\ge$ 30 K at the broad emissions by a C-type shock-heating model. \citet{riq10} revealed a high SiO to HCO$^{+}$ ratio in the foot point MC and \citet{riq18} concluded the enhancement of SiO caused by C-type shocks with 30--50 km~s$^{-1}$. \citet{tor10a} inferred that a magnetic-reconnection at the positions of the broad emissions can possibly be responsible for the source of shock-heating, although they could not explain why does it take place at the positions of the broad emissions instead of the bottom of the U shape.

Figures \ref{lbch1} and \ref{lbch2} in the Appendix show that all colliding clump clouds emit SiO. \citet{tsu15a} indicated that CCCs within $\sim$10$^{5}$ years is likely accompanied by SiO emissions originating from C-type shocks. Therefore, Our CCC scenario proposes the shock-heating model as originating from the CCC. This scenario naturally explains the origin of the high temperature and the enhancement of SiO in the foot point MC.
%%%%%%%%%%

\subsection{Comparison with other CCC regions}
We finally discuss the properties of CCCs in the GC. By comparing CCCs in the Galactic disk region to the GC, we found the following differences between them:
\begin{enumerate}
  \item CCCs in the GC have a few to ten times larger relative velocity between the two colliding clouds, and
  \item CCCs in the GC are not necessarily associated with high-mass stars or clusters. (In contrast, CCCs in the disk are always associated with high-mass stars or clusters)
\end{enumerate}
These two points may be linked. 

\citet{tak14} have performed hydrodynamic simulations of CCCs between non-identical clouds having Bonner--Ebert profiles. They discussed the importance of the relative velocity of the clouds in forming a massive clump and triggering high-mass star formation. In their simulations, they found that a larger relative velocity can enhance the formation of cores, which grow into a massive clump by coalescing with each other at the compressed layer. On the other hand, they also found that a excessive relative velocity suppresses the growth of a massive clump by hampering mass accretion due to the short accretion time available. They finally concluded that a relative velocity that is fast enough to allow significant core formation but slow enough to give these cores time to accrete is necessary for a CCC to trigger high-mass star formation.

To elucidate above hypothesis, we collected physical parameters of all CCC objects reported so far, especially for their relative velocities, column densities, and the number of high mass stars. Table~2 summarizes physical parameters of colliding clouds.
\begin{figure}[!htbp]
  \begin{center}
    \includegraphics[width=16.5cm]{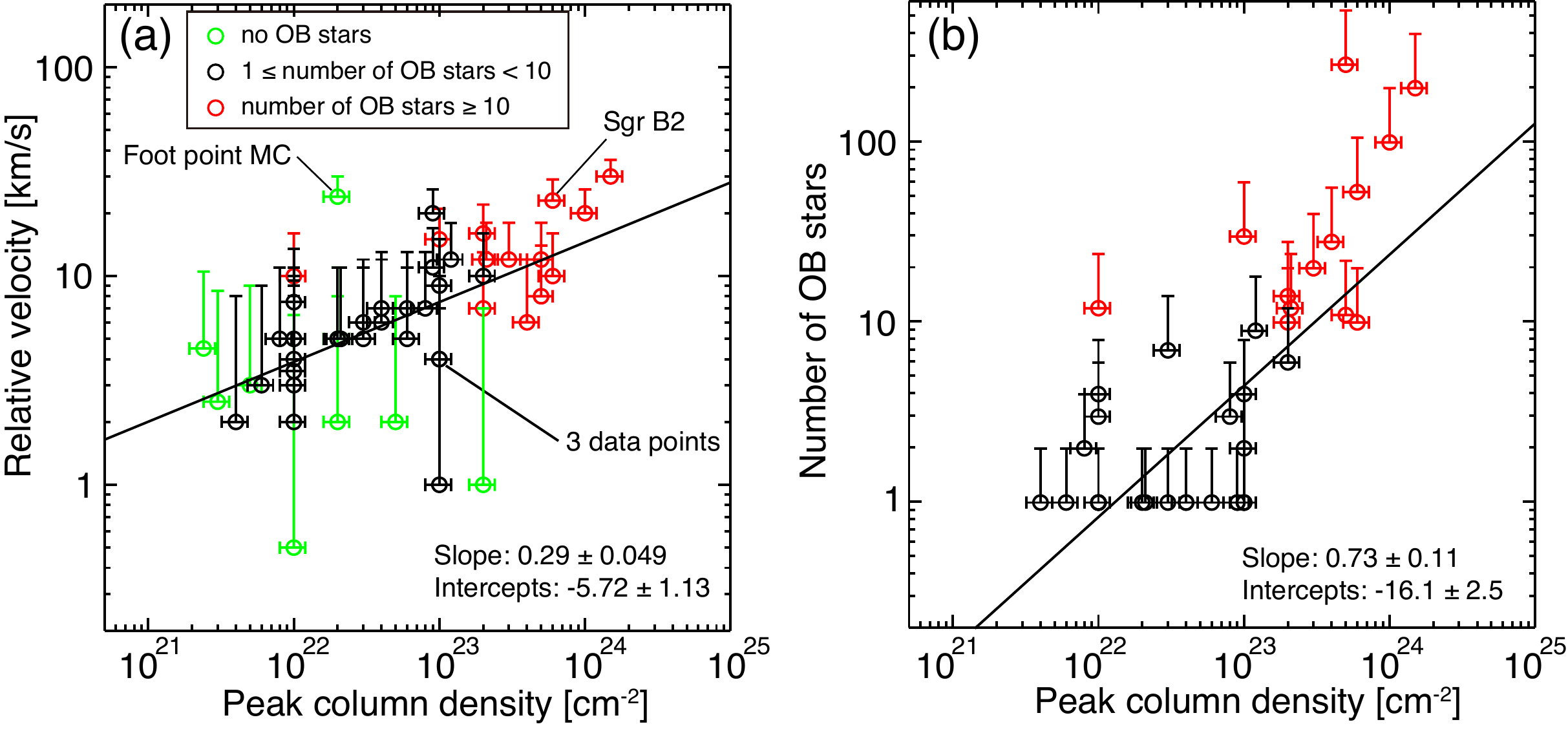}
  \end{center}
  \caption{(a) Scatter plot of the peak column density and relative velocity of colliding clouds of the Galactic sources on a double-logarithmic scale. The black, red, and light green symbols, respectively, indicate CCCs associated with clusters having less than ten O- and early B-type stars, more than ten O- and early B-type stars, and without O- and early B-type stars. The black line indicates the best-fit result of black and red symbols using a least-squares method. (b) Scatter plot of the peak column density and the number of O- and early B- type stars of colliding clouds of the Galactic sources on a double-logarithmic scale. The black line indicates the best-fit result of black and red symbols using a least-squares method.}\label{plot}
\end{figure}

Figure \ref{plot} show scatter plots between the peak column density and relative velocity (\ref{plot}a), and the peak column density and the number of OB stars (\ref{plot}b) of the CCCs in the Galactic sources reported to date. The difference in symbol color corresponds to the number of O- and early B-type stars triggered by the CCC. The column of column density in Table~2 includes both peak column densities and corrected peak column densities. We estimated the typical difference between a typical column density and a peak column density to be factor of $\sim$3 from our CO data (Table \ref{table:physpara}). The corrected peak column density was derived from multiplying the typical column density in reference(s) by 3. The error bars were determined as $\pm$20 $\%$ by assuming that the error is dominated by the intensity calibration error.
The measured relative velocities in literatures summarized in Table~2 is a line-of-sight relative velocity and true three dimensional relative velocity must be larger than the measured value. \citet{wil11} observed galactic disk clouds in some nearby galaxies by JCMT and measured a cloud-to-cloud-velocity-dispersion to be 6 km~s$^{-1}$. \citet{dob15} performed hydrodynamic simulations of a spiral galaxy and also confirmed that the dispersion is 4 -- 6 km~s$^{-1}$ in their simulations. Therefore, we here put +6 km~s$^{-1}$ ~as the errors of the relative velocity in the plot. To estimate the error of the number of O- and early B-type stars is difficult, however it should be less than factor of 2. Then, we adopt 2 as the error.

The CCCs observed so far in the disk presumably achieve these conditions.
The plot clearly exhibits a linear trend between the two parameters on the logarithmic axes, and the top-right region is dominated by CCCs with higher-mass clusters. This result implies that a collision with a higher relative velocity requires a higher column density to trigger star formation. This is compatible with the conclusion of \citet{tak14}.
The foot point MC, which shows clear signatures of CCCs but which is not associated with any high-mass star-formation activity, has a similar column density to that seen in CCCs in the Galactic disk, in spite of its larger relative velocity. This may be the first case in which the large relative velocity of the collision suppresses the formation of high-mass stars. In contrast, Sgr~B2 ---represented as an open blue rectangle in the figure--- is a famous star-burst region in the Milky Way, hosting $\sim$50 ultra-compact H$_\mathrm{II}$ regions \citep{gau95}, and also it is known to have originated as a CCC \citep{has94}. It is possible that this region has a large relative velocity between the two clouds but that, due to its high column density, an extreme CCC occurred and triggered the star burst. 

The collision timescale in the GC is shorter than for a CCC in the disk. This is attributed purely to the large relative velocity of the GC clouds. If the elapsed time exceeds $\sim$10$^6$ years, there is no way to recognize it to be a CCC in the GC. However, thanks to this short elapsed time and to little dispersal of the clouds by ionization due to UV radiation from high-mass stars, the (clump-hole) shapes of pairs of colliding clouds are preserved. Another property of the CCCs in the GC is enhanced SiO emission. Figures \ref{lbch1} and \ref{lbch2} in the Appendix show that all colliding clump clouds emit SiO. Indeed, SiO emission accompanies all the possible CCCs in the CMZ observed so far (\cite{has94, tsu15a,tsu15b}). This emission may be evidence for supersonic shock waves, which are generated at the beginning of a collision with a large relative velocity. Therefore, all the recognizable CCCs in the GC are accompanied by SiO emission.

The magnetic activities in the GC are strong enough to affect the gas dynamics \citep{eno14}. Although the detailed mechanism for the formation of a massive clump in such strong magnetic field is still unclear, an extreme CCC might take place at stagnation points of molecular gas, which may occur at the foot points of loops and/or at contact points between x1 and x2 orbits \citep{bin91}.

%*************** ABOVE: DISCUSSION

%----------------- FROM HERE: SUMMARY
\section{Summary}
 For the first time, we have applied the CCC-identification methodology developed by studies of CCCs in the Galactic disk to the GC. To examine the properties of the CCCs in the GC, as the first step we applied this method to the common foot point of molecular loops 1 and 2, where the systematic downflow may enhances the chance of CCCs occuring.
 We summarize the conclusions of the present work as follows:

\begin{enumerate}
 \item Two pairs of velocity components (main component 1--sub-component 1 and main component 2--sub-component 2) show complementary distributions in space and V shapes in velocity. These are evidence of CCCs.
 \item The CCCs took place 10$^{5-6}$ years ago, either by contact between loops 1 and 2 and/or by loops 1 and 2 to impacting clouds in the Galactic disk. Our new model, based on CCCs, completely explains the origin of the U shapes and the broad emissions in the foot point MC. This model is consistent with the kinematic models proposed in previous works.
 \item In contrarast to the Galactic disk, the CCCs in the foot point MC are not associated with any high-mass-star-formation activity. We suggest that the large relative velocity suppresses the formation of high-mass stars by the CCCs as proposed by \citet{tak14}.
 \end{enumerate}
%*************** ABOVE: SUMMARY

\begin{ack}
We thank (an) anonymous referee(s) for helpful comments that improved the manuscript.
The Mopra telescope is funded by the Commonwealth of Australia as a National Facility managed by CSIRO as part of the Australia Telescope. The UNSW-MOPS spectrometer used was funded by the Australian Research council with the support of the Universities of New South Wales, Sydney and Macquarie, together with the CSIRO. The ASTE project is driven by Nobeyama Radio Observatory (NRO), a division of National Astronomical Observatory of Japan (NAOJ), in collaboration with University of Chile, and Japanese institutes, including the University of Tokyo, Nagoya University, Osaka Prefecture University, Ibaraki University, and Hokkaido Univsersity. Observations with ASTE were in part carried out remotely from Japan by using NTT's GEMnet2 and its partnet R\&E (Research and Education) networks, which are based on AccessNova collaboration of University of Chile, NTT Laboratories, and NAOJ.
\end{ack}

\newpage

\begin{figure}[!htbp]
  \begin{center}
    \includegraphics[width=16.5cm]{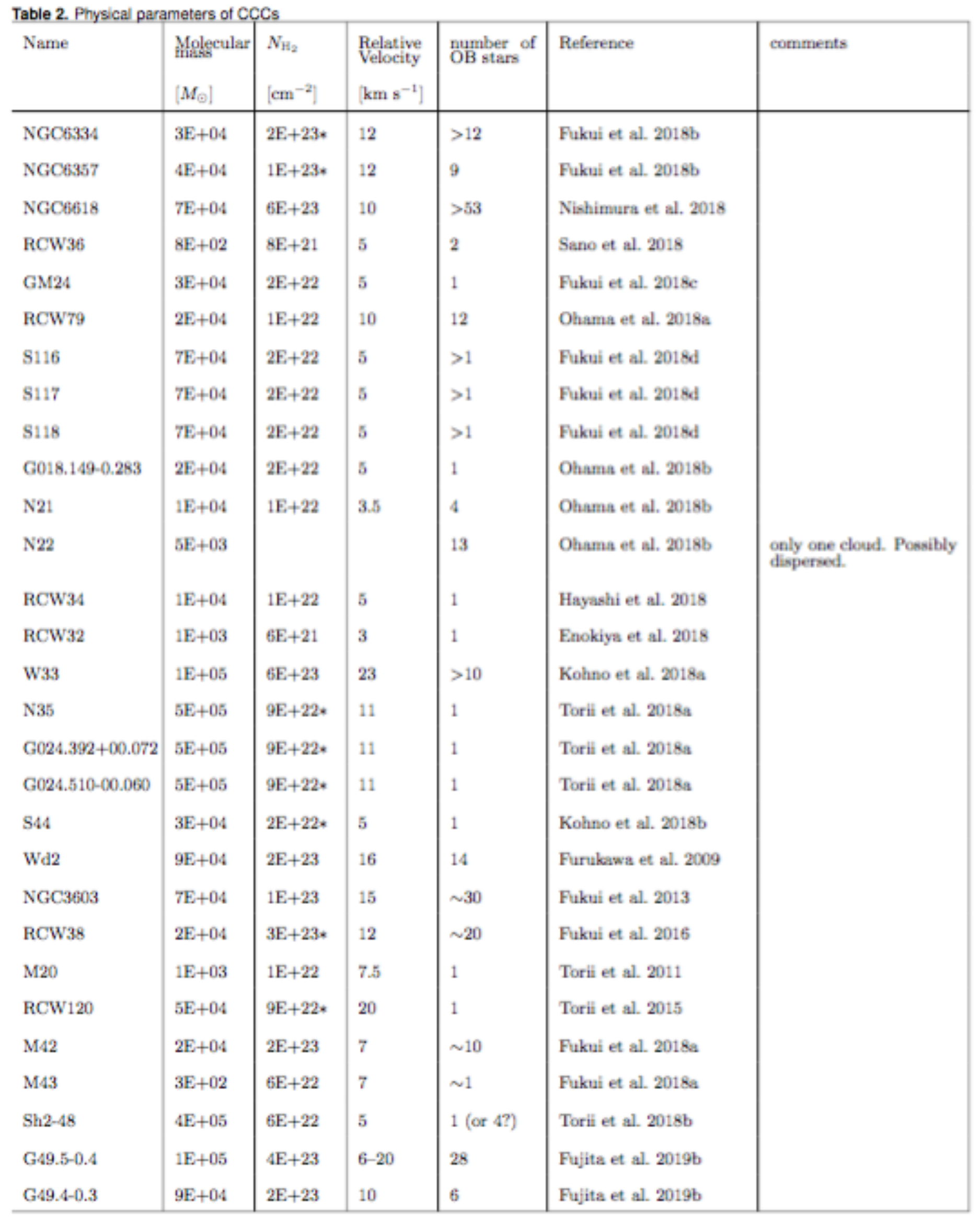}
  \end{center}
  \caption{}\label{}
\end{figure}

\begin{figure}[!htbp]
  \begin{center}
    \includegraphics[width=16.5cm]{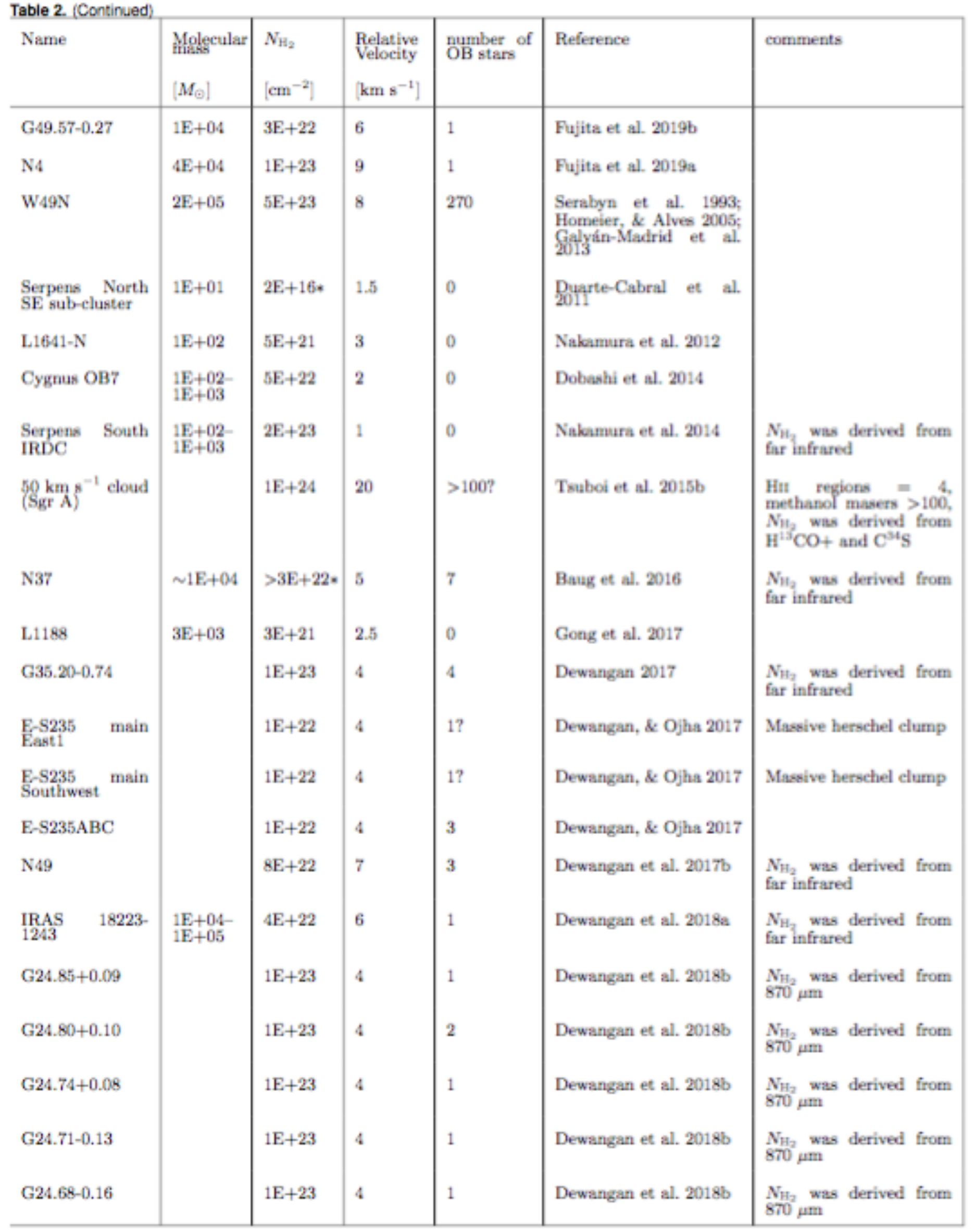}
  \end{center}
  \caption{}\label{}
\end{figure}

\begin{figure}[!htbp]
  \begin{center}
    \includegraphics[width=16.5cm]{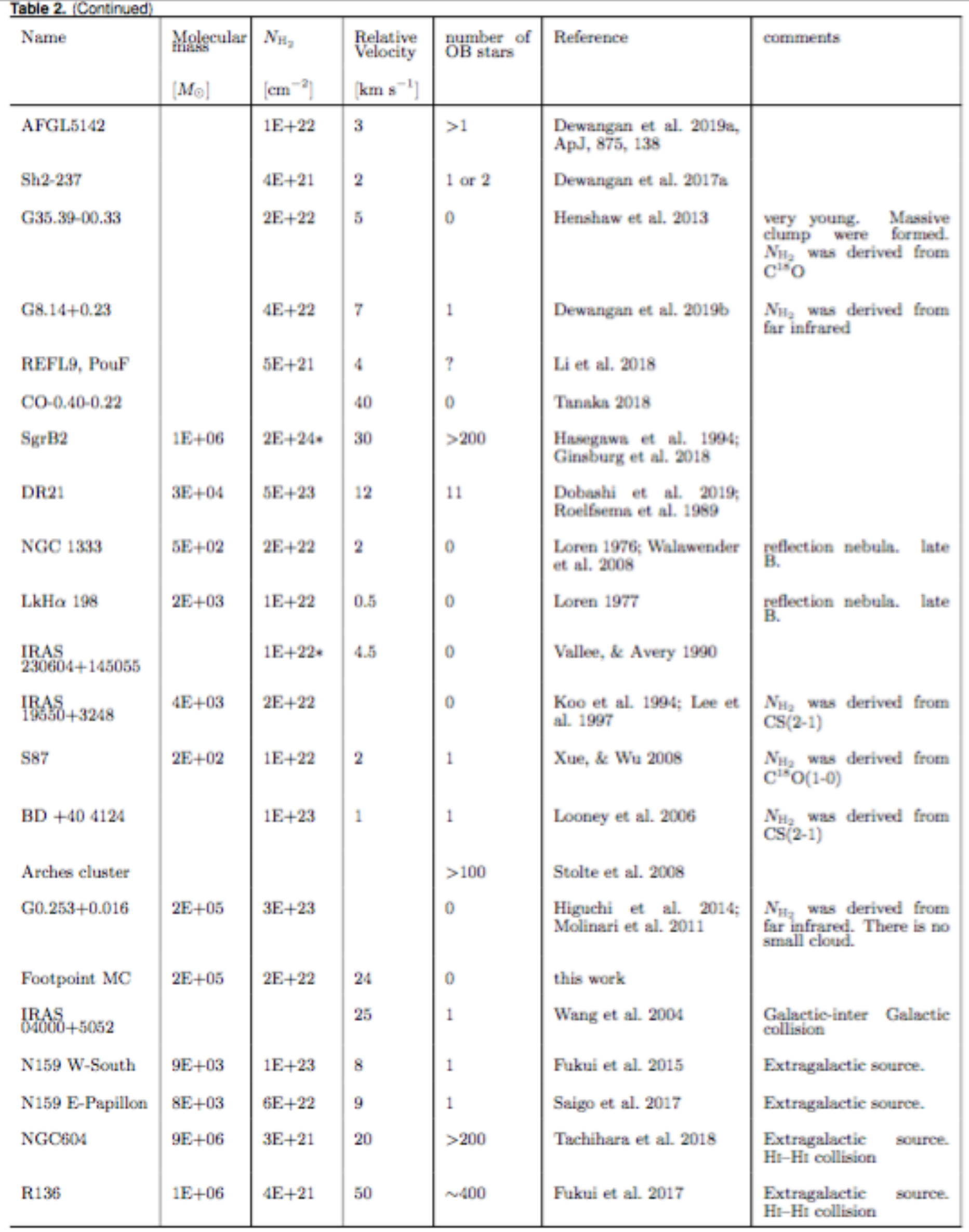}
  \end{center}
  \caption{}\label{}
\end{figure}

\begin{figure}[!htbp]
  \begin{center}
    \includegraphics[width=16.5cm]{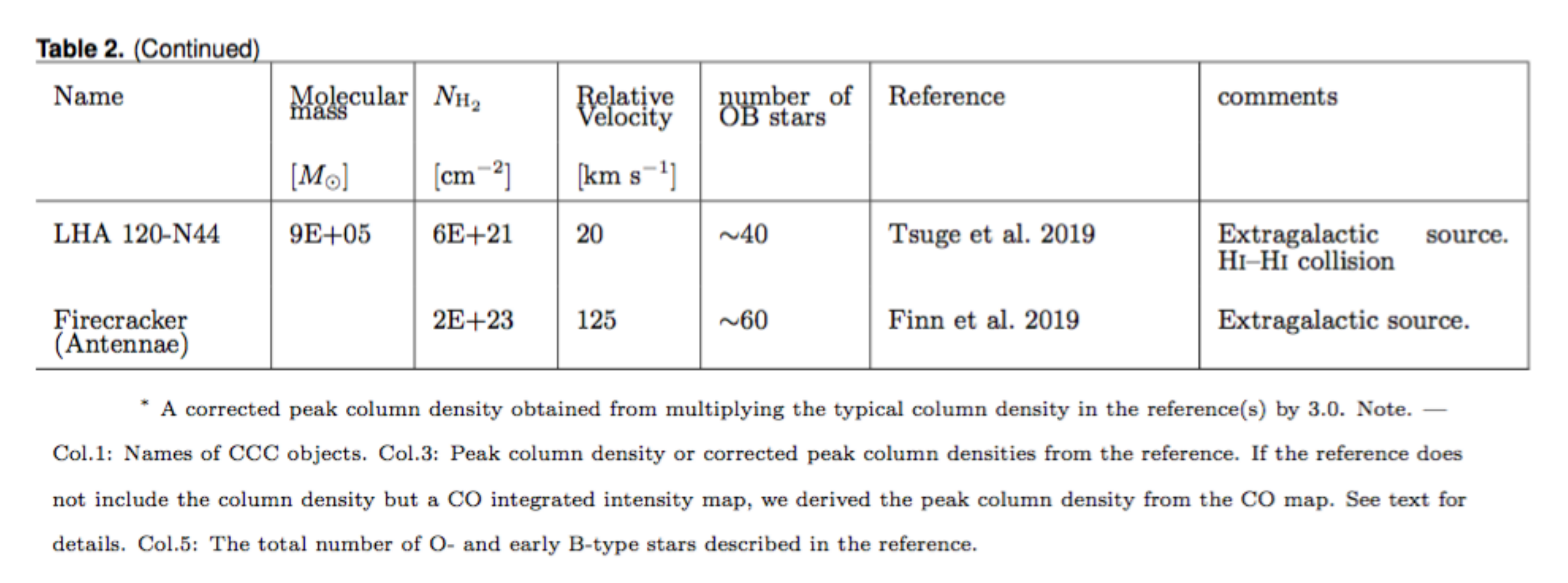}
  \end{center}
  \caption{}\label{}
\end{figure}

%----------------- FROM HERE: APPENDIX
\appendix 
\section*{Detailed Distribution of Molecular Clouds}
\begin{figure}[!htbp]
  \begin{center}
    \includegraphics[width=16.5cm]{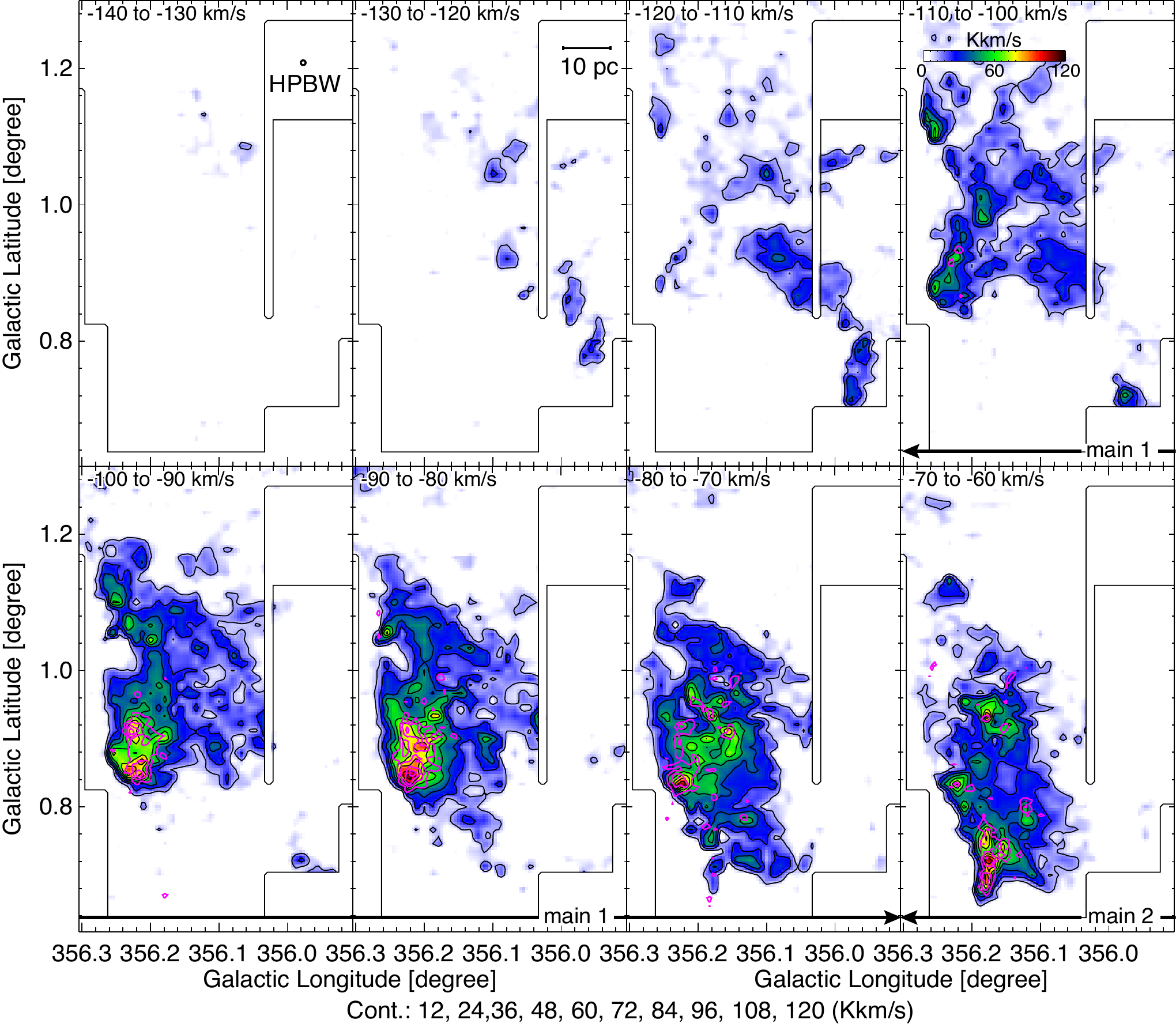}
  \end{center}
  \caption{Velocity channel distribution toward the foot point MC in $^{12}$CO($J$=3--2). The magenta contours show the distribution of SiO($J$=2--1) obtained by \citet{riq18}.}\label{lbch1}
\end{figure}

\begin{figure}[!htbp]
  \begin{center}
    \includegraphics[width=16.5cm]{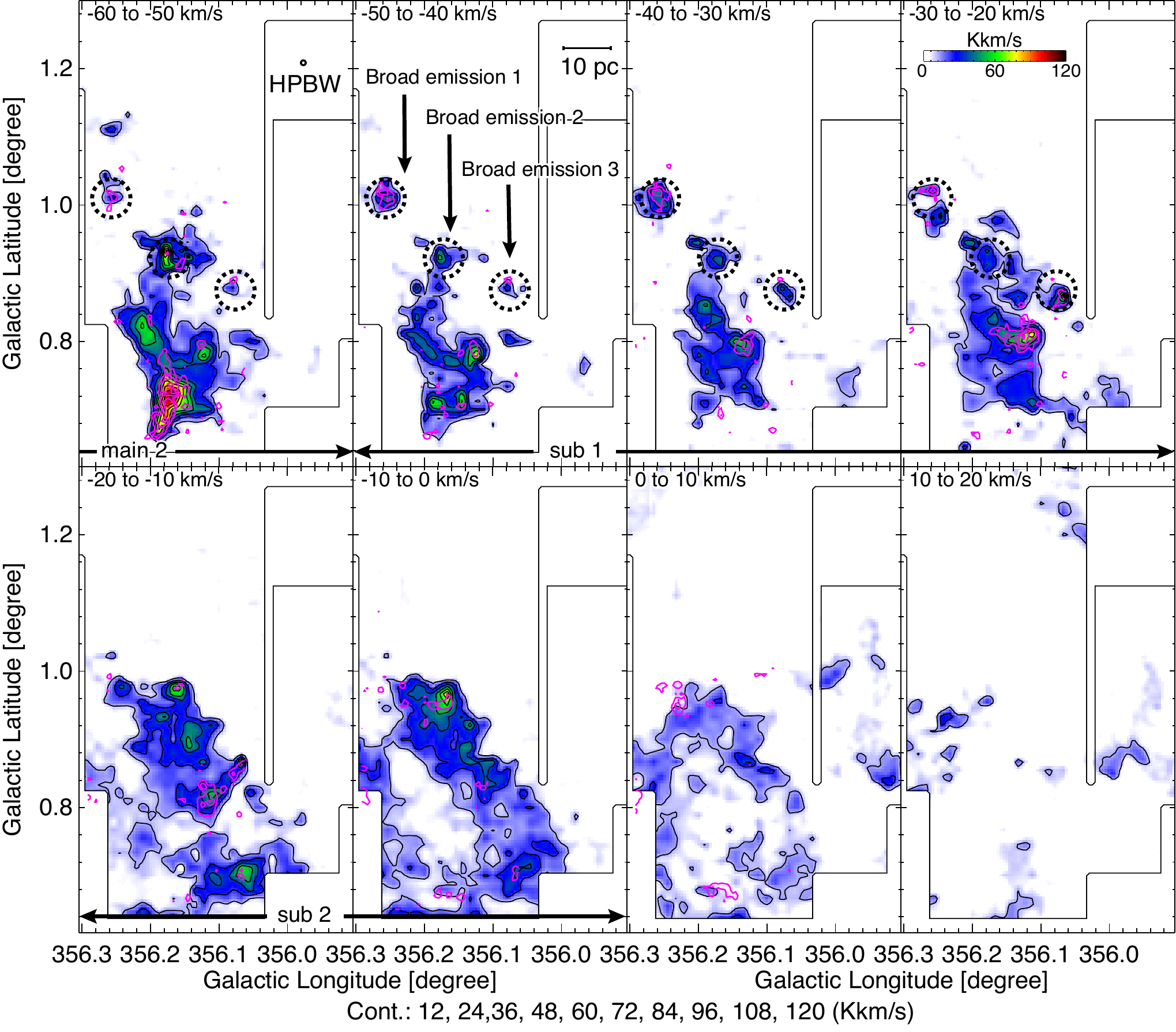}
  \end{center}
  \caption{Velocity channel distribution toward the foot point MC in $^{12}$CO($J$=3--2). The magenta contours show the distribution of SiO($J$=2--1) obtained by \citet{riq18}.}\label{lbch2}
\end{figure}

Figures \ref{lbch1} and \ref{lbch2} show the velocity-channel distributions of $^{12}$CO($J$=3--2) ~toward the foot point MC. The distribution of SiO($J$=2--1) is superposed on the $^{12}$CO($J$=3--2) ~as magenta contours.

%\section{Case of two or more paragraphs}
%*************** ABOVE: APPENDIX

%----------------- FROM HERE: REFERENCES
%%%
% See the manual for the detail.
%%%

%*************** ABOVE: REFERENCES


\begin{thebibliography}{}
% Journals(e.g. A\&A,ApJ,AJ,NMRAS,PASP ...)
% Authors, Year, Journal, Vol#, Page#
% Journal Title Abbreviation >> http://www.asj.or.jp/pasj/Jabb.html
\bibitem[Bania (1977)]{ban77}  Bania, T. M.\ 1977, \apj, 216, 381
\bibitem[Baug et al.(2016)]{bau16} Baug, T., Dewangan, L.~K., Ojha, D.~K., et al.\ 2016, \apj, 833, 85
\bibitem[Binney et al.(1991)]{bin91}  Binney, J., Gerhard, O. E., Stark, A. A., Bally, J., \& Uchida, K. I.\ 1991, \mnras, 252, 210
\bibitem[Bitran et al.(1997)]{bit97}  Bitran, M., Alvarez, H., Bronfman, L., May, J., \& Thaddeus, P.\ 1997, \aaps, 125, 99
\bibitem[Dewangan et al.(2017a)]{dew17a} Dewangan, L.~K., Ojha, D.~K., Zinchenko, I., et al.\ 2017a, \apj, 834, 22
\bibitem[Dewangan(2017)]{dew17} Dewangan, L.~K.\ 2017, \apj, 837, 44
\bibitem[Dewangan, \& Ojha(2017)]{dewo17} Dewangan, L.~K., \& Ojha, D.~K.\ 2017, \apj, 849, 65
\bibitem[Dewangan et al.(2017b)]{dew17b} Dewangan, L.~K., Ojha, D.~K., \& Zinchenko, I.\ 2017b, \apj, 851, 140
\bibitem[Dewangan et al.(2018a)]{dew18a} Dewangan, L.~K., Ojha, D.~K., Zinchenko, I., et al.\ 2018a, \apj, 861, 19
\bibitem[Dewangan et al.(2018b)]{dew18b} Dewangan, L.~K., Dhanya, J.~S., Ojha, D.~K., et al.\ 2018b, \apj, 866, 20
\bibitem[Dewangan et al.(2019a)]{dew19a} Dewangan, L.~K., Ojha, D.~K., Baug, T., et al.\ 2019a, \apj, 875, 138
\bibitem[Dewangan et al.(2019b)]{dew19b} Dewangan, L.~K., Sano, H., Enokiya, R., et al.\ 2019b, \apj, 878, 26
\bibitem[Dobashi et al.(2014)]{dob14} Dobashi, K., Matsumoto, T., Shimoikura, T., et al.\ 2014, \apj, 797, 58
\bibitem[Dobashi et al.(2019)]{dob19} Dobashi, K., Shimoikura, T., Katakura, S., et al.\ 2019, \pasj, 58
\bibitem[Dobbs et al.(2015)]{dob15} Dobbs, C.~L., Pringle, J.~E., \& Duarte-Cabral, A.\ 2015, \mnras, 446, 3608
\bibitem[Duarte-Cabral et al.(2011)]{dua11} Duarte-Cabral, A., Dobbs, C.~L., Peretto, N., et al.\ 2011, \aap, 528, A50
\bibitem[Enokiya et al.(2014)]{eno14}  Enokiya, R., et al.\ 2014, \apj, 780, 72
\bibitem[Enokiya et al.(2018)]{eno18}  Enokiya, R., et al.\ 2018, \pasj, 70, 49
\bibitem[Ezawa et al.(2008)]{eza08}  Ezawa, H., et al.\ 2008, \procspie, 7012, 701208
\bibitem[Ezawa et al.(2004)]{eza04}  Ezawa, H., Kawabe, R., Kohno, K., \& Yamamoto, S.\ 2004, \procspie, 5489, 763
\bibitem[Finn et al.(2019)]{fin19} Finn, M.~K., Johnson, K.~E., Brogan, C.~L., et al.\ 2019, \apj, 874, 120
\bibitem[Fujita et al.(2019a)]{fuj19a} Fujita, S., Torii, K., Tachihara, K., et al.\ 2019a, \apj, 872, 49
\bibitem[Fujita et al.(2019b)]{fuj19b} Fujita, S., Torii, K., Kuno, N., et al.\ 2019b, \pasj, 46
\bibitem[Fukui et al.(2006)]{fuk06}  Fukui, Y., et al.\ 2006, Science, 314, 106
\bibitem[Fukui et al.(2013)]{fuk13}  Fukui, Y., et al.\ 2013, \apj, 780, 36
\bibitem[Fukui et al.(2015)]{fuk15} Fukui, Y., Harada, R., Tokuda, K., et al.\ 2015, \apjl, 807, L4
\bibitem[Fukui et al.(2016)]{fuk16} Fukui, Y., Torii, K., Ohama, A., et al.\ 2016, \apj, 820, 26
\bibitem[Fukui et al.(2017)]{fuk17} Fukui, Y., Tsuge, K., Sano, H., et al.\ 2017, \pasj, 69, L5
\bibitem[Fukui et al.(2018a)]{fuk18a}  Fukui, Y., et al.\ 2018a, \apj, 859, 166
\bibitem[Fukui et al.(2018b)]{fuk18b} Fukui, Y., Kohno, M., Yokoyama, K., et al.\ 2018b, \pasj, 70, S41
\bibitem[Fukui et al.(2018c)]{fuk18c} Fukui, Y., Kohno, M., Yokoyama, K., et al.\ 2018c, \pasj, 70, S44
\bibitem[Fukui et al.(2018d)]{fuk18d} Fukui, Y., Ohama, A., Kohno, M., et al.\ 2018d, \pasj, 70, S46
\bibitem[Furukawa et al.(2009)]{fur09}  Furukawa et al.\ 2009, \apjl, 696, 115
\bibitem[Galv{\'a}n-Madrid et al.(2013)]{gal13} Galv{\'a}n-Madrid, R., Liu, H.~B., Zhang, Z.-Y., et al.\ 2013, \apj, 779, 121
\bibitem[Gaume et al.(1995)]{gau95}  Gaume, R. A., Claussen, M. J., de Pree, C. G., Goss, W. M., and Mehringer, D. M.\ 1995, \apj, 449, 663
\bibitem[Ginsburg et al.(2018)]{gin18} Ginsburg, A., Bally, J., Barnes, A., et al.\ 2018, \apj, 853, 171
\bibitem[Gong et al.(2017)]{gon17} Gong, Y., Fang, M., Mao, R., et al.\ 2017, \apjl, 835, L14
\bibitem[Habe \& Ohta (1992)]{hab92}  Habe, A., \& Ohta, K.\ 1992, \pasj, 44, 203
\bibitem[Hasegawa et al.(1994)]{has94}  Hasegawa, T., Sato, F., Whiteoak, J., B., and Miyawaki, R.\ 1994, \apj, 429, 77
\bibitem[Haworth et al.(2015)]{haw15}  Haworth et al.\ 2015, \mnras, 454, 1634
\bibitem[Hayashi et al.(2018)]{hay18} Hayashi, K., Sano, H., Enokiya, R., et al.\ 2018, \pasj, 70, S48
\bibitem[Henshaw et al.(2013)]{hen13} Henshaw, J.~D., Caselli, P., Fontani, F., et al.\ 2013, \mnras, 428, 3425
\bibitem[Higuchi et al.(2014)]{hig14} Higuchi, A.~E., Chibueze, J.~O., Habe, A., et al.\ 2014, \aj, 147, 141
\bibitem[Homeier, \& Alves(2005)]{hom05} Homeier, N.~L., \& Alves, J.\ 2005, \aap, 430, 481
\bibitem[Inoue \& Fukui (2013)]{ino13}  Inoue, T., \& Fukui, Y.\ 2013, \apjl, 774, 31
\bibitem[Kakiuchi et al.(2018)]{kak18}  Kakiuchi, K., et al.\ 2018, \mnras, 476, 5629
\bibitem[Kaneda et al.(2012)]{kan12}  Kaneda, H., et al.\ 2012, \pasj, 64, 25
\bibitem[Kohno (2005)]{koh05}  Kohno, K.\ 2005, \asp, 344, 242
\bibitem[Kohno et al.(2018a)]{koh18a} Kohno, M., Torii, K., Tachihara, K., et al.\ 2018a, \pasj, 70, S50
\bibitem[Kohno et al.(2018b)]{koh18b} Kohno, M., Tachihara, K., Fujita, S., et al.\ 2018b, \pasj, 126
\bibitem[Koo et al.(1994)]{koo94} Koo, B.-C., Lee, Y., Fuller, G.~A., et al.\ 1994, \apj, 429, 233
\bibitem[Kudo et al.(2011)]{kud11}  Kudo, N., et al.\ 2011, \pasj, 63, 171
\bibitem[Lee et al.(1997)]{lee97} Lee, H.-G., Koo, B.-C., Park, Y.-S., et al.\ 1997, \pasj, 49, 639
\bibitem[Li et al.(2018)]{li18} Li, C., Wang, H., Zhang, M., et al.\ 2018, \apjs, 238, 10
\bibitem[Looney et al.(2006)]{loo06} Looney, L.~W., Wang, S., Hamidouche, M., et al.\ 2006, \apj, 642, 330
\bibitem[Loren(1976)]{lor76} Loren, R.~B.\ 1976, \apj, 209, 466
\bibitem[Loren(1977)]{lor77} Loren, R.~B.\ 1977, \apj, 218, 716
\bibitem[Machida et al.(2009)]{mac09}  Machida, M., et al.\ 2009, \pasj, 61, 411
\bibitem[Matsumoto et al.(1988)]{mat88} Matsumoto, R., Horiuchi, T., Shibata, K., et al.\ 1988, \pasj, 40, 171
\bibitem[Molinari et al.(2011)]{mol11} Molinari, S., Bally, J., Noriega-Crespo, A., et al.\ 2011, \apjl, 735, L33
\bibitem[Morris \& Serabyn (1996)]{mor96}  Morris, M., \& Serabyn, E.\ 1996, \araa, 34, 645
\bibitem[Nakamura et al.(2012)]{nak12} Nakamura, F., Miura, T., Kitamura, Y., et al.\ 2012, \apj, 746, 25
\bibitem[Nakamura et al.(2014)]{nak14} Nakamura, F., Sugitani, K., Tanaka, T., et al.\ 2014, \apjl, 791, L23
\bibitem[Nishimura et al.(2018)]{nis18} Nishimura, A., Minamidani, T., Umemoto, T., et al.\ 2018, \pasj, 70, S42
\bibitem[Ohama et al.(2018a)]{oha18a} Ohama, A., Kohno, M., Hasegawa, K., et al.\ 2018a, \pasj, 70, S45
\bibitem[Ohama et al.(2018b)]{oha18b} Ohama, A., Kohno, M., Fujita, S., et al.\ 2018b, \pasj, 70, S47
\bibitem[Oka et al.(1998)]{oka98}  Oka, T., Hasegawa, T., Sato, F., Tsuboi, M., \& Miyazaki, A.\ 1998, \apjs, 118, 455
\bibitem[Riquelme et al.(2010)]{riq10}  Riquelme, D., Bronfman, L., Mauersberger, R., May, J., \& Wilson, T. L.\ 2010, \aap, 523, 45
\bibitem[Riquelme et al.(2018)]{riq18}  Riquelme, D., et al.\ 2018, \aap, 613, 42  
\bibitem[Roelfsema et al.(1989)]{roe89} Roelfsema, P.~R., Goss, W.~M., \& Geballe, T.~R.\ 1989, \aap, 222, 247
\bibitem[Parker (1966)]{par66}  Parker, E. N.\ 1966, \apj, 145, 811
\bibitem[Saigo et al.(2017)]{sai17} Saigo, K., Onishi, T., Nayak, O., et al.\ 2017, \apj, 835, 108
\bibitem[Sano et al.(2018)]{san18} Sano, H., Enokiya, R., Hayashi, K., et al.\ 2018, \pasj, 70, S43
\bibitem[Serabyn et al.(1993)]{ser93} Serabyn, E., Guesten, R., \& Schulz, A.\ 1993, \apj, 413, 571
\bibitem[Stolte et al.(2008)]{sto08} Stolte, A., Ghez, A.~M., Morris, M., et al.\ 2008, \apj, 675, 1278
\bibitem[Suzuki et al.(2015)]{suz15}Suzuki, T., Fukui, Y., Torii, K., Machida, M., \& Matsumoto, R.\ 2015, \mnras, 454, 3049
\bibitem[Tachihara et al.(2018)]{tac18} Tachihara, K., Gratier, P., Sano, H., et al.\ 2018, \pasj, 70, S52
\bibitem[Takahashi et al.(2009)]{tak09}Takahashi, K., et al.\ 2009, \pasj, 61, 957
\bibitem[Takahira et al.(2014)]{tak14}Takahira, K., Tasker, E.~J., and Habe, A.\ 2014, \apj, 792, 63
\bibitem[Takekawa et al.(2019)]{tak19} Takekawa, S., Oka, T., Iwata, Y., Tsujimoto, S., \& Nomura, M.\ 2019, \apj, 871, 1
\bibitem[Tanaka(2018)]{tan18} Tanaka, K.\ 2018, \apj, 859, 86
\bibitem[Torii et al.(2010a)]{tor10a}  Torii, K., et al.\ 2010a, \pasj, 62, 675
\bibitem[Torii et al.(2010b)]{tor10b}  Torii, K., et al.\ 2010b, \pasj, 62, 1307
\bibitem[Torii et al.(2011)]{tor11}  Torii, K., et al.\ 2011, \apj, 738, 46
\bibitem[Torii et al.(2015)]{tor15} Torii, K., Hasegawa, K., Hattori, Y., et al.\ 2015, \apj, 806, 7
\bibitem[Torii et al.(2018a)]{tor18a} Torii, K., Fujita, S., Matsuo, M., et al.\ 2018a, \pasj, 70, S51
\bibitem[Torii et al.(2018b)]{tor18b} Torii, K., Hattori, Y., Matsuo, M., et al.\ 2018b, \pasj, 121
\bibitem[Tsuboi et al.(2015a)]{tsu15a}Tsuboi, M., Miyazaki, A., and Uehara, K.\ 2015a, \pasj, 67, 90
\bibitem[Tsuboi et al.(2015b)]{tsu15b}Tsuboi, M., Miyazaki, A., and Uehara, K.\ 2015b, \pasj, 67, 109
\bibitem[Tsuge et al.(2019)]{tsu19} Tsuge, K., Sano, H., Tachihara, K., et al.\ 2019, \apj, 871, 44
\bibitem[Vallee, \& Avery(1990)]{val90} Vallee, J.~P., \& Avery, L.~W.\ 1990, \aap, 233, 553
\bibitem[Walawender et al.(2008)]{wal08} Walawender, J., Bally, J., Francesco, J.~D., et al.\ 2008, Handbook of Star Forming Regions, Volume I, 346
\bibitem[Wang et al.(2004)]{wan04} Wang, J.-J., Chen, W.-P., Miller, M., et al.\ 2004, \apjl, 614, L105
\bibitem[Wilson et al.(2000)]{wil00} Wilson, C.~D., Scoville, N., Madden, S.~C., et al.\ 2000, \apj, 542, 120.
\bibitem[Wilson et al.(2011)]{wil11} Wilson, C.~D., Warren, B.~E., Irwin, J., et al.\ 2011, \mnras, 410, 1409
\bibitem[Xue, \& Wu(2008)]{xue08} Xue, R., \& Wu, Y.\ 2008, \apj, 680, 446
\end{thebibliography}
\end{document}